\documentclass[sigconf]{aamas}  
\usepackage{balance}  

\usepackage{booktabs}

\usepackage{caption}
\renewcommand{\arraystretch}{1.5}
\usepackage{mwe}
\usepackage{tabularx}
\usepackage{enumitem}
\setlist[itemize]{leftmargin=*}
\usepackage{soul}
\usepackage{algorithm}
\usepackage{algpseudocode}
\usepackage[utf8]{inputenc}
\usepackage[free-standing-units]{siunitx} 
\usepackage{bm}
\usepackage{balance}
\usepackage{listings}
\usepackage{multirow}
\usepackage{pdfpages}
\usepackage{balance}
\usepackage{setspace}
\usepackage[framemethod=tikz]{mdframed}
\usepackage[justification=centering]{caption}
\usepackage{hyperref}
\usepackage{subcaption}
\usepackage{graphicx}
\graphicspath{ {images/} }
\usepackage{afterpage}
\usepackage{float}
\usepackage{url}
\usepackage{dirtytalk}
\usepackage{amsmath}
\usepackage[export]{adjustbox}
\usepackage{amssymb}
\DeclareMathOperator*{\argmax}{argmax} 
\newcommand{\R}{\mathbb{R}}
\usepackage{array}
\newcolumntype{P}[1]{>{\centering\arraybackslash}p{#1}}
\usepackage{tabu}
\usepackage{makecell}
\definecolor{black}{rgb}		{0.0, 0.0, 0.0}
\definecolor{white}{rgb}		{1.0, 1.0, 1.0}
\definecolor{yellow}{rgb}		{1.0, 1.0, 0.8}
\definecolor{red}{rgb}			{0.6, 0.0, 0.2}
\definecolor{blue}{rgb}		{0.0, 0.2, 0.5}
\definecolor{green}{rgb}		{0.6, 0.8, 0.8}
\definecolor{dark_green}{RGB} {0, 140, 0}
\definecolor{gold}{rgb}		{0.6, 0.4, 0.1}
\definecolor{grey}{RGB}{0,0,0}
\definecolor{Gray}{gray}{0.8}
\definecolor{MediumGray}{gray}{0.9}
\definecolor{LightGray}{gray}{0.98}
\definecolor{LightCyan}{rgb}{0.88,1,1}
\definecolor{purple}{RGB}{128,0,128}
\definecolor{sl_blue}{RGB}{47, 60, 105}
\definecolor{orange}{RGB}{255,165,0}
\definecolor{Gray}{gray}{0.85}

\usepackage{xcolor,colortbl}

\newcolumntype{a}{>{\columncolor{grey}}c}
\newcolumntype{b}{>{\columncolor{white}}c}

\setcopyright{ifaamas}  
\acmDOI{}  
\acmISBN{}  
\acmConference[AAMAS'19]{Proc.\@ of the 18th International Conference on Autonomous Agents and Multiagent Systems (AAMAS 2019)}{May 13--17, 2019}{Montreal, Canada}{N.~Agmon, M.~E.~Taylor, E.~Elkind, M.~Veloso (eds.)}  
\acmYear{2019}  
\copyrightyear{2019}  
\acmPrice{}  

\begin{document}

\title{Negative Update Intervals in Deep \\ Multi-Agent Reinforcement Learning} 


\author{Gregory Palmer}
\affiliation{%
  \institution{University of Liverpool, UK}
}
\email{G.J.Palmer@liverpool.ac.uk}

\author{Rahul Savani}
\affiliation{%
  \institution{University of Liverpool, UK}
}
\email{Rahul.Savani@liverpool.ac.uk}

\author{Karl Tuyls}
\affiliation{%
  \institution{University of Liverpool, UK}
}
\email{K.Tuyls@liverpool.ac.uk}

\begin{abstract}  
In Multi-Agent Reinforcement Learning (MA-RL), independent cooperative learners must overcome a number of pathologies to learn optimal joint policies. Addressing one pathology often leaves approaches vulnerable towards others. For instance, \emph{hysteretic Q-learning} \cite{matignon2007hysteretic} addresses miscoordination while leaving agents vulnerable towards misleading stochastic rewards. Other methods, such as \emph{leniency}, have proven more robust when dealing with multiple pathologies simultaneously \cite{JMLR:v17:15-417}. However, leniency has predominately been studied within the context of strategic form games (bimatrix games) and fully observable Markov games consisting of a small number of probabilistic state transitions. This raises the question of whether these findings scale to more complex domains. For this purpose we implement a temporally extend version of the \emph{Climb Game} \cite{claus1998dynamics}, within which agents must overcome multiple pathologies \emph{simultaneously}, including relative overgeneralisation, stochasticity, the alter-exploration and moving target problems, while learning from a large observation space. We find that existing lenient and hysteretic approaches fail to consistently learn near optimal joint-policies in this environment. To address these pathologies we introduce \emph{Negative Update Intervals-DDQN (NUI-DDQN)}, a Deep MA-RL algorithm which discards episodes yielding cumulative rewards outside the range of expanding intervals. NUI-DDQN consistently gravitates towards optimal joint-policies in our environment, overcoming the outlined pathologies.
\end{abstract}

%


\maketitle


\section{Introduction}

The \emph{Multi-Agent Reinforcement Learning} (MA-RL) literature provides a rich taxonomy of learning pathologies that cooperative \emph{Independent Learners} (ILs) must overcome to converge upon an optimal joint-policy, e.g. \emph{stochasticity}, the \emph{alter-exploration} and \emph{moving target} problems \cite{matignon2012independent}. While searching for an optimal joint-policy, the actions of ILs influence each others' search space. This can lead to \emph{action shadowing}, where miscoordination due to sub-optimal joint-policies results in utility values of optimal actions being underestimated \cite{fulda2007predicting, panait2006lenience}. In this paper we address the above pathologies and a type of action shadowing called \emph{relative overgeneralisation}. This pathology can occur when pairing an IL's available actions with arbitrary actions by the other agents results in a sub-optimal action having the highest utility estimate \cite{JMLR:v17:15-417}. As a result, ILs can be drawn to sub-optimal but wide peaks in the reward search space due to a greater likelihood of achieving collaboration there \cite{panait2006lenience}. 

Numerous methods have been proposed in MA-RL literature to help ILs cope with the outlined pathologies. Traditionally methods have been studied within the context of bimatrix games and fully observable Markov games consisting of a small number of probabilistic state transitions. However, finding robust solutions that perform consistently is challenging, even in traditional settings, as solutions to one pathology often leave agents vulnerable towards others \cite{JMLR:v17:15-417}. For approaches that perform consistently, questions remain regarding scalability, i.e., can they overcome the same pathologies in complex domains that suffer from the curse of dimensionality and require reasoning over long time horizons? To answer this question we evaluate the ability of \emph{leniency} and \emph{hysteretic Q-Learning}, two decentralized approaches with a strong track record in traditional settings, to overcome the pathologies outlined in a temporally extended, partially observable version of the \emph{Climb Game} \cite{claus1998dynamics}. We call this game the \emph{Apprentice Firemen Game} (AFG). 

Hysteretic Q-Learning is a form of optimistic learning that helps agents overcome miscoordination \cite{matignon2007hysteretic}. However, this approach can leave agents vulnerable towards misleading stochastic rewards \cite{JMLR:v17:15-417,palmer2018lenient}. Lenient learners meanwhile are initially forgiving towards teammates, often ignoring state transitions that would lower a utility value \cite{panait2006lenience}. However, the more often an observation-action pair is encountered, the less likely lenient learners are to be forgiving. Therefore leniency is less vulnerable towards misleading stochastic rewards \cite{JMLR:v17:15-417}. Both approaches have been extended to \emph{Deep} MA-RL (MA-DRL) \cite{omidshafiei2017deep,palmer2018lenient}. To date the majority of MA-DRL research focuses on stochasticity and mitigating an amplified moving target problem resulting from obsolete state transitions being stored inside \emph{Experience Replay Memories} (ERM) \cite{foerster2017stabilising,palmer2018lenient,zheng2018weighted,omidshafiei2017deep}. The AFG meanwhile allows us to study the robustness of MA-DRL algorithms simultaneously facing all of the above pathologies in a system suffering from the curse of dimensionality.

Within the AFG agents must make an irrevocable decision that will determine the outcome of an episode. We find that while hysteretic and lenient learners deliver promising performances in layouts where agents can observe each other's irrevocable decision, both algorithms converge upon sub-optimal joint policies when the same \emph{irrevocable} decision is made in \emph{seclusion}. To help ILs overcome the outlined pathologies in this challenging setting, we introduce a novel approach where agents maintain expanding intervals estimating the $min$ and $max$ of cumulative reward distributions for state-transition trajectories ending without miscoordination. The intervals determine which trajectories are stored and used for sampling, allowing ILs to discard trajectories resulting in miscoordination. This reduces the impact of noisy utility values occurring in cooperative games with high punishment for uncoordinated behavior, increasing the likelihood of average utility values being established for sequences of actions leading to coordinated outcomes. We call this approach NUI-DDQN (\emph{Negative Update Intervals Double-DQN}).

Our main contributions can be summarized as follows. \\
\noindent{\textbf{1)}} We design a new environment that simultaneously confronts ILs with all four of the mentioned pathologies. The environment is based on the Climb Game, which has been used to study relative overgeneralisation and stochastic rewards. We embed the Climb Game in a temporally-extended gridworld setting, that we call the \emph{Apprentice Firemen Game} (AFG), in which two fireman need to coordinate to extinguish a fire. Stochastic transitions can be added through introducing randomly moving civilians who obstruct paths.\\ 
\noindent{\textbf{2)}} We empirically evaluate hysteretic and lenient approaches in two \emph{AFG} layouts (Figure \ref{fig:env}). \emph{Layout 1} examines whether the pathologies can be overcome when ILs can observe each other while making an irrevocable choice that determines the outcome of an episode. In contrast \emph{layout 2} requires ILs to independently make the same \emph{irrevocable} decision in \emph{seclusion}. We find that ILs predominately converge upon superior joint-policies in \emph{layout 1}, providing evidence that ILs can implicitly learn to avoid miscoordination when able to observe each other during transitions that determine an episode's outcome. \emph{Layout 2} poses a challenge for existing approaches. Lenient learners in particular face the following dilemma: remain lenient and be led astray by misleading stochastic rewards, or estimate average utility values and succumb to relative overgeneralisation due oscillating utility values caused by stochastic transitions.\\
\noindent{\textbf{3)}} We introduce \emph{NUI-DDQN}, a MA-DRL algorithm which discards episodes yielding cumulative rewards outside the range of expanding intervals. These intervals are maintained for sequences of transitions (trajectories) equivalent to actions from \emph{cooperative games}. NUI-DDQN reduces the noise introduced by punishing values resulting from miscoordination to utility estimates, allowing ILs to overcome \emph{relative overgeneralisation} and the \emph{alter-exploration problem}. \emph{NUI-DDQN} consistently converges upon the optimal joint-policy in both layouts for deterministic and stochastic rewards.


\section{Definitions and Notations}

Below is a summary of the definitions used in this paper. Although many of the concepts discussed are from game theory we shall use the terms \emph{agent} and \emph{player} interchangeably. Furthermore, to prevent confusion we refer to actions taken in strategic-form games as $u \in U$ while actions in Markov games are denoted as $a \in A$.

\noindent{\bf Strategic-Form Games.} A strategic-form game is defined as a tuple $(n, U_{1...n}, R_{1...n})$ where $n$ represents the number of agents, $U_{1...n}$ the joint action space $(U_1 \times ... \times U_n)$, with $U_i$ being the set of actions available to agent $i$, and $R_i$ is the reward function $R_i : U_1 \times ... \times U_n \rightarrow \R$ for each agent $i$ \cite{bowling2002multiagent}. In this paper, we focus on strategic-form games with $n=2$, commonly known as bimatrix games.
We note that in strategic-form games players make their choices~\emph{simultaneously}.

\noindent{\bf Stochastic rewards.} Normally in a strategic-form game, all rewards are deterministic (DET). Given that stochasticity (of both rewards and transitions) are a central pathology of MA-RL, in this paper we also consider both \emph{partially stochastic} (PS) and \emph{fully stochastic} (FS) rewards. DET and FS reward functions exclusively return deterministic and stochastic rewards respectively for each $U_{1...n}$. Meanwhile for a PS reward function there exists up to $|U|-1$ joint actions $U_{1...n}$ for which a stochastic reward is returned, while the remaining joint actions return deterministic rewards~\cite{JMLR:v17:15-417}. 


\noindent{\bf Markov Games.} 
In contrast to strategic-form games where choices are made simultaneously, in a Markov game players make their choices \emph{sequentially}. The game transitions from state to state, with choices and corresponding rewards collected along the way. Formally, a Markov game $M$ has a finite state space $X$, an observation function $O_i : X \rightarrow \R^d$ which returns a $d$-dimensional observation for agent $i$, for each state $x \in X$ a joint action space $(A_1 \times ... \times A_n)$, with $A_i$ being the number of actions available to agent $i$, a transition function $T : X \times A_1 \times ... \times A_n \times X' \rightarrow [0,1]$, returning the probability of transitioning from a state $x$ to $x'$ given an action profile $a_1 \times ... \times a_n$, and a reward function: $R_i : X \times  A_1 \times ... \times A_n \rightarrow \R$ for each agent $i$ \cite{shapley1953stochastic,leibo2017multi}. We allow \emph{terminal states} at which the game ends. Note that a strategic-form game can be thought of a Markov game with a single terminal state.

\noindent{\bf Policies.} For agent $i$, the policy $\pi_i$ represents a mapping from the observation space to a probability distribution over actions: $\pi_i : O_i \rightarrow \Delta(A_i)$, while $\bm{\pi}$ refers to a joint policy of all agents. Joint policies excluding agent $i$ are defined as $\bm{\pi}_{-i}$. The notation $\langle \pi_i, \bm{\pi}_{-i}\rangle$ refers to a joint policy with agent $i$ following $\pi_i$ while the remaining agents follow $\bm{\pi}_{-i}$. 

\noindent{\bf Trajectories.} We call
a particular roll-out of a policy $\pi_i$, i.e., the sequence of resulting states, actions, and associated rewards, a \emph{trajectory} and denote it by~$\tau_i$~\cite{sutton1999between}. 

\noindent{\bf Expected Gain.} Given a joint policy $\bm{\pi}$ the gain (or expected sum of future rewards) for each agent $i$ starting from a state $x$ is defined in Equation \ref{eq:gain} below, where $r_{i,t}$ refers to the reward received by agent $i$ at time-step $t$, while $\gamma \in [0,1)$ is a discount factor \cite{matignon2012independent}: 
\begin{equation} \label{eq:gain}
\mathcal{U}_{i,\bm{\pi}}(x) = E_\pi \Bigl\{ \sum_{k=0}^\infty \gamma^k r_{i,t+k+1} | x_t = x \Bigr\}.
\end{equation}

\noindent{\bf Nash Equilibrium.} For a Markov game, a joint policy $\bm{\pi}^*$ is a \emph{Nash equilibrium} if and only if no agent $i$ can improve it's gain through unilaterally deviating from $\bm{\pi}^*$ \cite{matignon2012independent}:
\begin{equation} \label{eq:nash_equilibrium}
\forall i, \forall \pi_i \in \Delta(X,A_i), \forall x \in X, \mathcal{U}_{i, \langle \pi^*_i,\bm{\pi}^*_{-i}\rangle}(x) \geq  \mathcal{U}_{i, \langle \pi_i,\bm{\pi}^*_{-i}\rangle}(x).
\end{equation}

\noindent{\bf Pareto Optimality.} From a group perspective Nash equilibria are often sub-optimal. In contrast Pareto-optimality defines a joint policy $\bm{\hat{\pi}}$ from which no agent $i$ can deviate without making at least one other agent worse off. A joint policy $\bm{\pi}$ is therefore Pareto-dominated by $\bm{\hat{\pi}}$ \emph{iff} \cite{matignon2012independent}:
\begin{equation} \label{eq:pareto_optimal}
\resizebox{0.9\hsize}{!}{%
$\forall i, \forall x \in X, \mathcal{U}_{i,\bm{\hat{\pi}}}(x) \geq \mathcal{U}_{i,\bm{\pi}}(x) \text{ and } \exists j, \exists x \in X, \mathcal{U}_{j,\bm{\hat{\pi}}}(x) \geq \mathcal{U}_{j,\bm{\pi}}(x)$}.
\end{equation}

\noindent{\bf Pareto Dominated Nash Equilibrium.} A joint policy $\bm{\hat{\pi}}^*$ is Pareto optimal if it is not Pareto-dominated by any other $\bm{\pi}$ \cite{matignon2012independent}. 

\noindent{\bf Team Games.} A game, be it a strategic-form or Markov game, is a \emph{team game} if every player gets the same reward, i.e., $R_1 = R_2 = R$. Thus, team games are \emph{fully cooperative} settings, where players have a shared objective to maximize their common return (see Figure \ref{fig:climb_game_versions}) \cite{claus1998dynamics,bucsoniu2010multi, matignon2012independent}.
In the rest of the paper, we only consider team games.

\section{\bf Temporally-extended Team Games} 

We conduct empirical evaluations in \emph{temporally-extended versions of team bimatrix games} (TEGs). The outcome of these games is determined by joint-trajectories $\bm{\tau}$ resulting from $\bm{\pi}$. The reward function has inequalities mirroring those of the corresponding bimatrix game. Therefore, each $\tau_i$ belongs to a set of trajectories $T^u$ that implements an action $u \in U$. ILs are tasked with learning a joint policy $\bm{\pi}$ that results in optimal joint trajectories $\bm{\tau}$. Throughout this paper $\tau_i$ refers to a trajectory that consists of all the state-transition tuples $(o_{t-1}, a_{t-1}, r_t, o_{t})$ of an individual episode. In Section~\ref{sec:AFG}, we introduce a temporally-extended version of the Climb Game, which serves as the basis for our experiments.

\section{Pathologies in MA-RL} \label{sec:Pathologies} 


In this section, we elaborate on the pathologies outlined in the introduction. As mentioned, addressing one pathology often leaves agents vulnerable towards others. We discuss this in detail, while considering the implications of tackling these pathologies in complex environments. To understand the pathologies, we consider two types of ILs attempting to estimate the quality of an action $u$ when paired with the actions $U'$ available to the other agent~\cite{JMLR:v17:15-417}: 

\begin{itemize}
    \item \emph{Average based learners} estimate the quality of $u$ based on the average return: $quality(u) = \sum_{u' \in U'} R_i(u,u')/|U'|$.
    \item \emph{Maximum based learners} estimate the quality of $u$ based on the maximum return observed: $quality(u) = \max_{u' \in U'} R_i(u,u')$. 
\end{itemize}

We now discuss the four pathologies: relative overgeneralisation, stochasticity,  the alter-exploration, and moving target problems.

\subsection{Relative Overgeneralisation} \label{sec:relative_overgeneralisation}

Relative overgeneralisation is a type of \emph{Action Shadowing}, occurring in games where a sub-optimal Nash Equilibrium yields a higher payoff on average when each selected action is paired with an arbitrary action chosen by the other player \cite{JMLR:v17:15-417}. A \emph{shadowed equilibrium} is an equilibrium defined by a policy $\overline{\bm{\pi}}$ that is shadowed by a policy $\bm{\hat{\pi}}$ in a state $x$, where at least one agent exists who when unilaterally deviating from $\overline{\pi}$, will receive a gain $U_{\langle \pi_i, \overline{\pi}_{-i} \rangle}(x)$ less than the minimum gain that can be obtained for deviating from $\bm{\hat{\pi}}$ (Equation \ref{eq:shadowed_eq}) \cite{matignon2012independent}. Relative overgeneralisation occurs in games where, as a result of a \emph{shadowed equilibrium}, the agents converge upon a sub-optimal Nash Equilibrium that is Pareto-dominated by at least one other Nash Equilibrium \cite{JMLR:v17:15-417,matignon2012independent,edselc.2-52.0-4154912397120080301}. 

\begin{equation} \label{eq:shadowed_eq}
\exists i \exists \pi_i  \mathcal{U}_{\langle \pi_i, \bm{\overline{\pi}}_{-i} \rangle}(x) < \min_{j, \pi_j}  \mathcal{U}_{\langle \pi_j, \bm{\hat{\pi}}_{-j} \rangle}(x) .
\end{equation}

\noindent{\bf Example:} Variations of the Climb Game \cite{claus1998dynamics} (Figure~\ref{fig:climb_game_versions}) are frequently used to study the susceptibility of ILs towards relative overgeneralisation. In the Climb Game the Pareto-Optimal Nash Equilibrium is $(A, A)$. However, assuming two ILs initially choose each of the actions available with equal probability, using an \emph{average based} algorithm, then \emph{Agent 1} will estimate that $C$ should be preferred over $A$ and $B$, since $\sum (A, j) < \sum (C, j)$ and $\sum (B, j) < \sum (C, j)$ for each of \emph{Agent 2's} actions $j$ \cite{JMLR:v17:15-417}. \emph{Agent 2} will come to the same conclusion, resulting in the agents gravitating towards the \emph{shadow equilibrium} $(C, C)$. If an alternative action is still being played with a small probability, then \emph{Agent 1} will move from action $C$ to $B$. Subsequently \emph{Agent 2} will also \textit{climb} from $C$ to $B$. At this point the agents will climb no further, having reached a Pareto dominated sub-optimal Nash equilibrium~$(B, B)$. 
\subsection{Stochasticity of rewards and transitions} \label{sec:Stochasticity}

In the deterministic reward Climb Game (Figure \ref{fig:climb_game_payoff_matrix}) \emph{relative overgeneralisation} can be overcome with \emph{maximum-based learning}, where agents consider each action $i$ based on the observed $max_j(i,j)$ \cite{JMLR:v17:15-417}. However, this approach leaves agents vulnerable towards misleading stochastic rewards. For example, in the \emph{Partially Stochastic Climb Game} (Figure \ref{fig:ps_climb_game_payoff_matrix}) the joint action $(B,B)$ yields stochastic rewards of 14 and 0 with 50\% probability. Therefore maximum based learners are drawn towards $(B,B)$, despite each agent only receiving a reward of 7 on average. In temporally extended games additional stochasticity can emerge as a result of environmental factors such as noisy observations and probabilistic state transitions. Meanwhile, ILs facing the curse of dimensionality must overcome challenges introduced by noisy approximated utility estimates backed-up from stochastic follow-on state-transitions or rewards \cite{matignon2012independent}.
\vspace{-3mm}
\begin{figure}[h]
\begin{subfigure}[b]{0.47\columnwidth}
\centering
\resizebox{\columnwidth}{!}{%
\def\arraystretch{1.2}%
\begin{tabular}{cc|c|c|c|}
  & \multicolumn{1}{c}{} & \multicolumn{3}{c}{Player 2}\\
  & \multicolumn{1}{c}{} & \multicolumn{1}{c}{$A$}  & \multicolumn{1}{c}{$B$} & \multicolumn{1}{c}{$C$} \\\cline{3-5}
  \multirow{3}*{\rotatebox{90}{Player 1}}  & $A$ & $(11,11)$ & $(-30,-30)$ & $(0,0)$ \\\cline{3-5}
  & $B$ & $(-30,-30)$ &  $(7,7)$ & $(6,6)$ \\\cline{3-5}
  & $C$ & $(0,0)$ & $(0,0)$ &  $(5,5)$ \\\cline{3-5}
\end{tabular}}
\caption{Deterministic}
\label{fig:climb_game_payoff_matrix}
\end{subfigure} 
\hspace{2mm}
\begin{subfigure}[b]{0.47\columnwidth}
\centering
\resizebox{\columnwidth}{!}{%
\def\arraystretch{1.2}%
\begin{tabular}{cc|c|c|c|}
  & \multicolumn{1}{c}{} & \multicolumn{3}{c}{Player 2}\\
  & \multicolumn{1}{c}{} & \multicolumn{1}{c}{$A$}  & \multicolumn{1}{c}{$B$} & \multicolumn{1}{c}{$C$} \\\cline{3-5}
  \multirow{3}*{\rotatebox{90}{Player 1}}  & $A$ & $(11,11)$ & $(-30,-30)$ & $(0,0)$ \\\cline{3-5}
  & $B$ & $(-30,-30)$ &  $(14/0,14/0)$ & $(6,6)$ \\\cline{3-5}
  & $C$ & $(0,0)$ & $(0,0)$ &  $(5,5)$ \\\cline{3-5}
\end{tabular}}
\caption{Partially Stochastic.}
\label{fig:ps_climb_game_payoff_matrix}
\end{subfigure} 
\caption{Climb Games Variations. For (b) joint-action $\bm{(B,B)}$ yields stochastic rewards of 14 and 0 with 50\% probability}
\label{fig:climb_game_versions}
\end{figure} 
\vspace{-6mm}

\subsection{The alter-exploration problem} \label{sec:alter_exploration}

The exploration-exploitation trade-off required by reinforcement learners adds to the challenge of learning noise-free utility estimates. Matignon et al. \cite{matignon2012independent} define global exploration, the probability of at least one of $n$ agents exploring as $1 - (1 - \epsilon)^n$, where each agent explores according to a probability $\epsilon$. In environments with a \emph{shadowed equilibrium}, as defined in Section \ref{sec:relative_overgeneralisation}, higher global exploration can result in agents converging upon a sub-optimal joint policy, as exploration can lead to penalties \cite{matignon2012independent}. Furthermore, we consider that in TEGs agents face two types of alter-exploration:\\ 
\noindent{\bf 1.) Exploring at an atomic level.} Due to global exploration short sequences of actions may be sub-optimal, e.g., collisions with other agents, sub-optimal paths, etc, thereby introducing noise when the agents compute utility value estimates.\\ 
\noindent{\bf 2.) Exploring at a trajectory level.} Global exploration can impact the set of trajectories $T^u$ that trajectory $\tau_i$ belongs to for an agent $i$, and therefore the outcome of an episode. We therefore argue that for MA-DRL algorithms to overcome the alter-exploration problem considerations are required at both levels. 
 
\subsection{The moving target problem}  \label{sec:moving_target}

When multiple ILs update policies in parallel an environment can no longer be considered Markovian, thereby losing the property that guarantees convergence for a large number of single-agent learning algorithms \cite{bowling2002multiagent,sutton1998introduction}. This problem is amplified in \emph{MA-DRL}, where using ERMs often results in deprecated transitions being sampled during training \cite{foerster2017stabilising,omidshafiei2017deep,palmer2018lenient}. Furthermore, as we shall discuss in section \ref{sec:Impact_of_stochastic_transitions}, stochastic transitions can lead to the moving target problem, which can result in long periods of miscoordination.

\section{Independent Learner Baselines}

The algorithms evaluated are extensions of the \emph{Double-DQN} (DDQN) introduced by Van Hasselt et al. \cite{van2016deep}. Each agent $i$ is implemented with a \emph{ConvNet} trained to approximate Q-Values for observation-action pairs: $Q_i: O_i \times A_i \rightarrow \R$ \cite{leibo2017multi}. The network parameters $\theta$ are trained using Adam \cite{kingma2014adam} on the mean squared Bellman residual with the expectation taken over state transitions uniformly sampled from an ERM based on a probability distribution $p\left(o,a\right)$ 
\cite{mnih2015human,lin1992self},
\begin{equation}\label{eq:time_dependent_loss_function}
L_i\left(\theta_i\right) = \mathbf{E}_{o,a \sim p\left(\cdot\right)}\Big[ \left(Y_t - Q\left(o,a; \theta_t\right)\right)^2 \Big],
\end{equation}
where $Y_t$ is the target:
\begin{equation} \label{eq:DQNTarget}
Y_t \equiv r_{t+1} + \gamma Q(o_{t+1}, \argmax_{a \in A} Q(o_{t+1}, a; \theta_t);\theta_{t}').
\end{equation}

The set of parameters $\theta_{t}'$ in Equation \ref{eq:DQNTarget} belong to a more stable \emph{target network}, which is synchronised with the current network every $n$ transitions \cite{van2016deep}. Each IL agent $i$ is implemented with a separate ERM used to store state transitions as tuples $(o_{i,t-1}, a_{i,t-1}, r_{i,t}, o_{i,t})$, consisting of an observation $o_{i,t-1}$, action $a_{i,t-1}$, the resulting observation $o_{i,t}$ and the immediate reward $r_{i,t}$. To ensure obsolete transitions are eventually discarded ERMs are implemented as First-In First-Out (FIFO) queues \cite{leibo2017multi}.

\subsection{Hysteretic Q-Learning} \label{sec:hyst}

Hysteretic Q-Learning is an optimistic MA-RL algorithm originally introduced to address \emph{maximum based} learner's vulnerability towards stochasticity by using two learning rates $\alpha$ and $\beta$, where $\beta < \alpha$ \cite{JMLR:v17:15-417}. Given a TD-Error $\delta$, where $\delta = Y_t - Q\left(o_t,a_t; \theta_t\right)$, a hysteretic Q-value update is performed as described in Equation~\ref{eq:QValueUpdate}, where $\beta$ reduces the impact of negative Q-Value updates while learning rate $\alpha$ is used for positive updates \cite{matignon2007hysteretic}. However, \emph{hysteretic Q-Learners} still have a tendency to gravitate towards sub-optimal policies when receiving misleading stochastic rewards~\cite{JMLR:v17:15-417,palmer2018lenient}. 

\begin{equation}\label{eq:QValueUpdate}
  Q\left(x_t, a_t\right) = 
  \begin{cases}
    Q\left(x_t, a_t\right) + \alpha \delta &\text{if $\delta > 0$}.\\
    Q\left(x_t, a_t\right) + \beta \delta &\text{Otherwise}.
  \end{cases}
\end{equation}


\subsection{Leniency} Lenient learners have proven robust towards stochastic rewards by initially forgiving (ignoring) sub-optimal actions by teammates, while over time applying an \emph{average based} approach for frequently visited observation-action pairs \cite{panait2006lenience,panait2008theoretical,zheng2018weighted,JMLR:v17:15-417,palmer2018lenient}. The frequency with which negative updates are performed is determined by Equation \ref{eq:leniency} and a random variable $\chi \sim U (0, 1)$, with negative updates only taking place if $\chi > l(o_i, a_i)$ (Equation \ref{eq:lenientQValueUpdate}). Constant $K$ is a leniency moderation factor determining how the temperature value affects the drop-off in lenience \cite{JMLR:v17:15-417}. The temperature $\mathcal{T}_t(o_i, a_i)$ is decayed each time an $(o_i,a_i)$ pair is encountered. Meanwhile, Lenient-DDQNs (LDDQNs) store the leniency value computed at time $t$ inside the ERM, subsequently determining the frequency with which the corresponding transition can induce negative updates \cite{palmer2018lenient}.  

\begin{equation} \label{eq:leniency}
l(o_i, a_i) = 1 - e^{-K*\mathcal{T}_t(o_i,a_i)}.
\end{equation}
\begin{equation}\label{eq:lenientQValueUpdate}
  Q\left(o_t, a_t\right) = 
  \begin{cases}
    Q\left(o_t, a_t\right) + \alpha \delta &\text{if $\delta > 0$ or $\chi > l\left(o_t, a_t\right)$}.\\
    Q\left(o_t, a_t\right) &\text{if $\delta \leq 0$ and $\chi \leq l\left(o_t, a_t\right)$}.
  \end{cases}
\end{equation}


\section{Negative~Update~Intervals~in~MA-DRL} \label{sec:NUIDDQN}

Our aim is to compute intervals for each action $u \in U$, where the lower endpoint is approximately the $min$ reward received for coordinated behaviour involving $u$. Therefore, receiving a reward less than $min$ on the interval indicates miscoordination has occurred. Below we define \emph{negative update intervals} and describe how they help NUI-DDQN overcome relative overgeneralisation in TEGs.

\noindent{\bf Negative Updates.} We define negative updates as Q-value updates using a TD-Error $\delta < 0$. In Equations \ref{eq:QValueUpdate} and \ref{eq:lenientQValueUpdate} both leniency and \emph{hysteretic Q-learning} reduce the impact of updates that would result in the lowering of Q-Values. However, neither algorithm distinguishes lowering Q-Values due to stochastic rewards from lowering based on miscoordination. Maximum based learners meanwhile avoid negative updates altogether by maintaining utility values for each action $u \in U$ based on the highest observed reward $r_u^{max}$.

\noindent{\bf Negative Update Intervals.} While agents guided by $r_u^{max}$ are vulnerable towards stochastic rewards, we consider that for PS and FS reward spaces where $r_u^{min}$ for coordinated outcomes is greater than the punishment received for miscoordination, there exists intervals $[r_u^{min},r_u^{max}]$ within which negative updates to utility estimates can be performed while mitigating the noise induced through punishment for miscoordination. We show that maintaining negative update intervals for each action $u \in U$ increases the likelihood of agents within TEGs converging upon an optimal joint policy $\bm{\hat{\pi}}^*$.

\noindent{\bf Classifying Trajectories.} Joint-trajectories $\bm{\tau}$ determine the rewards yielded by TEGs. Therefore, given an oracle $\vartheta : T \rightarrow U$ capable of determining the set $T^u$ that trajectory $\tau$ belongs to, negative update intervals $[r_u^{min},r_u^{max}]$ can be stored for each action $u \in U$, thereby increasing the likelihood of ILs computing noise free \emph{average} utility values for transitions belonging to coordinated joint-trajectories $\bm{\tau}$. For simplicity our evaluations use TEGs with a predefined $U$. However, under Future Work (Section \ref{ref:future_work:improved_oracle}) we discuss the potential of using a \emph{theory of mind neural network} \cite{pmlr-v80-rabinowitz18a} for $\vartheta$.

\noindent{\bf Maintaining Negative Update Intervals.} We establish $r_u^{max}$ for each action $u$ during an initial exploration phase used to fill the ERMs. Initially $r_u^{min} = r_u^{max}$. During training $r_u^{min}$ is gradually decayed. To prevent a premature decay during phases where ILs are confronted with the alter-exploration problem, 
we only only decay if the cumulative reward for the trajectory ($R^\tau = \sum_{t=0}^{|\tau|} r_t$) is large enough.
That is, $r_u^{min}$ is only decayed if $R^\tau \geq r_u^{max} - \varepsilon$, where $\varepsilon$ is a small constant.\\
\noindent{\bf Addressing Catastrophic Forgetting.} Catastrophic forgetting occurs when trained networks forget how to perform previously learned tasks while learning to master a new task \cite{goodfellow2013empirical}. To allow agents to maintain Q-Values for transitions belonging to less frequently observed actions $u$, without preventing outdated transitions from being discarded, we implement a separate ERM for each action $u \in U$. Instead of storing $n$ transitions each $ERM_u$ stores $n$ episodes, since traditional ERMs may store a significant number of obsolete transitions once ILs become efficient at solving a task and require less steps. Episodic ERMs meanwhile are more likely to reflect the current search space. During sampling the $ERM_u$ are concatenated.\\
\noindent{\bf Storing Trajectories.} In addition to $r_u^{min}$ we maintain vectors $R^u$, which store the most recent $n$ cumulative rewards for each action $u$. A trajectory is stored \emph{iff} the cumulative reward $R^\tau$ is greater than the \emph{max} between $r_u^{min}$ and the $R^u$'s mean $\overline{R^u}$ minus the standard deviation $SD_{R^u}$ (Equation \ref{eq:NUI_TRAJECTORY_DISCARD}). Therefore, while leniency is vulnerable towards miscoordination upon cooling temperature values, NUI-DDQN will continue to discard miscoordination trajectories. 

\begin{equation} \label{eq:NUI_TRAJECTORY_DISCARD}
ERM_u = \begin{cases}
    ERM_u \cup \tau &\text{if } R^\tau \geq max(r_u^{min}, \overline{R^u}-SD_{R^u}).\\
    ERM_u & \text{Otherwise}.
  \end{cases}
\end{equation}

\section{The Apprentice Firemen Game (AFG)}
\label{sec:AFG}

The Climb Game is often studied as a repeated game. We are interested in solving an equivalent game extended over the temporal dimension, where joint trajectories $\bm{\tau}$ result in outcomes comparable to the joint-actions from Figure \ref{fig:climb_game_versions}. We formulate a TEG based on the Climb Game that we call the \emph{Apprentice Firemen Game (AFG)}, where two (or more) agents located within a gridworld are tasked with locating and extinguishing fires. First however the agents must locate an equipment \emph{pickup area} and choose one of the items listed in Table \ref{tab:equipment} below. The task is fully cooperative, i.e. both agents are required to extinguish one fire. As outlined in Table \ref{tab:equipment} both agents detonating an explosive device (fighting fire with fire) is the most effective combination, equivalent to the joint action $(A, A)$ in the Climb Game. While the fire extinguisher is more effective than the fire blanket, agents choosing one run the risk of being hit by debris if the other agent triggers an explosive device, whereas the fire blanket offers protection. Therefore the fire extinguisher and fire blanket are equivalent to actions $B$ and $C$ respectively. 

\begin{figure}[h]
\centering
    \begin{subfigure}[b]{0.48\columnwidth}
        \centering
        \includegraphics[height=2.7cm, frame]{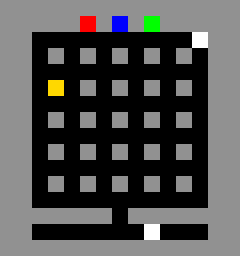} 
        \caption{Layout 1}
        \label{fig:env:shared_tool_pick}
    \end{subfigure}
    \begin{subfigure}[b]{0.48\columnwidth}
        \centering
        \includegraphics[height=2.7cm, frame]{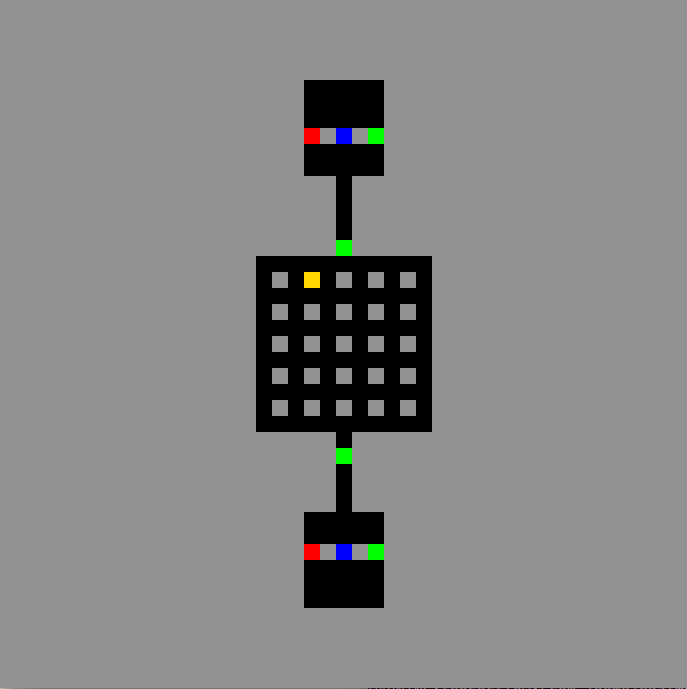} 
        \caption{Layout 2}
        \label{fig:env:city}
    \end{subfigure}
\caption{AFG layouts with fires (yellow), obstacles (grey) and equipment A (red), B (green) \& C (blue). Firemen are initially white, but following a pickup adopt the equipment's color.}
\label{fig:env}
\end{figure}

\def\arraystretch{1.1}%
\begin{table}[h]
\centering
\resizebox{0.8\columnwidth}{!}{%
\begin{tabular}{||c | c | c | c||}
\hline
\textbf{Description} & \textbf{Action ($\bm{u \in U}$)} & \textbf{Effectiveness} & \textbf{Risk} \\
\hline\hline
Explosive Device & A & High & High \\
\hline
Fire Extinguisher & B & Medium & High \\
\hline
Fire Blanket & C & Weak & None \\
\hline
\end{tabular}} 
\caption{Apprentice Firemen Game Equipment}
\label{tab:equipment}
\end{table}
\vspace{-2em}
ILs are not explicitly told which actions other agents have performed \cite{JMLR:v17:15-417}. However, we hypothesize ILs can learn to avoid miscoordination in the AFG when able to observe each other during transitions that determine an episode's outcome, reducing the impact of optimal joint action $(A, A)$ being a \emph{shadowed equilibrium}. To test this hypothesis we conduct experiments using two layouts outlined below (and illustrated in Figure \ref{fig:env}), where equipment pickup decisions are irrevocable for the duration of each episode. At the start of an episode one randomly chosen obstacle in the main area is set on fire. Episodes end when both agents occupy cells next to the fire, upon which a terminal reward is returned. To eliminate confounding factors all non-terminal transitions yield a reward of 0. We introduce a 10,000 step limit upon observing that trained agents delay miscoordination outcomes through avoiding the fire. The agents receive a miscoordination reward of -1.0 upon reaching this limit. The action space is discrete and includes moving up, down, left, right and NOOP. Pickup actions occur automatically upon ILs entering an equipment cell empty handed. DDQNs perform well when receiving rewards within $[-1, 1]$, which led us to choose the reward structures listed in Figure \ref{fig:Apprentice_Firemen_Games_Rewards}. For stochastic transitions randomly moving civilians can be added who obstruct paths.

\begin{algorithm}[h]
\caption{NUI-DDQN}
\label{alg:NUI-DDQN}
  \begin{algorithmic}[1]
    \State \textbf{Input:} Number of episodes $E$, replay period $K$, max steps $T$
    \State \textbf{Random exploration phase} (Init for $ERM_u$, $r_u^{min}$ and $r_u^{max}$) 
    \For{$e=1$ to $E$}
        \State $\tau=\emptyset$ 
        \State Observe $o_0$ and choose $a_0 \sim\pi_\theta(o_0)$
        \For{$t=1$ to $T$}
            \State Observe $o_t$, $r_t$
            \State Store transition ($o_{t-1}$, $a_{t-1}$, $r_{t}$, $o_{t}$) in $\tau$
            \If{$t \equiv 0 \mod K$}
                \State Optimise Network
            \EndIf
            \State Copy weights from time to time: $\theta_{t}' \gets \theta_t$
            \State Choose $a_t \sim\pi_\theta(o_t)$ 
        \EndFor    
        \State $u \gets $ $\vartheta(\tau)$
        \State $R^u \gets R^u \cup R^\tau$
        \If {$ r_u^{max} < R^\tau$}
            \State $r_u^{max} \gets R^\tau$
        \EndIf
        \If {$R^\tau \geq max(r_u^{min}, \overline{R^u}-SD_{R^u})$}
            \State $ERM_u \gets ERM_u \cup \tau$
        \EndIf 
        \If {$R^\tau \geq r_u^{max} - \varepsilon$}
            \State $r_u^{min} \gets decay(r_u^{min})$
        \EndIf
    \EndFor
    \end{algorithmic}
\end{algorithm}
\medskip
\noindent
\textbf{Layout 1: Observable irrevocable decisions} 
Two agents in a $16 \times 15$ gridworld begin each episode in opposite corners of a compartment separated from the main area. The agents must exit the compartment, gather equipment from a shared \emph{pickup area} and subsequently extinguish the fire, meaning that agents observe each other during the irrevocable equipment selection process. One agent can therefore observe the other agent's choice and subsequently select a best response to avoid miscoordination - in terms of the original Climb (bimatrix) game, this allows agents to act as if it was a perfect-information \emph{commitment} version of the game with a follower and leader. 

\medskip
\noindent
\textbf{Layout 2: Irrevocable decisions in seclusion}
Two agents in a padded $53 \times 53$ gridworld begin each episode in separate chambers. To mimic the simultaneity of the choice of actions in the Climb (bimatrix) game, each agent is limited to $13 \times 13$ centered observations. Agents are therefore unable to observe each others' equipment selection actions.  

\vspace{-2em}

\begin{figure*}
\centering
\resizebox{\linewidth}{!}{%
\def\arraystretch{1.1}%
 \begin{tabular}{||c | P{2.1cm} | P{2.1cm} | P{2.1cm}| | P{2.1cm} | P{2.1cm} | P{2.1cm}| | P{2.1cm} | P{2.1cm} | P{2.1cm}||} 
\hline     
    &
    \multicolumn{3}{|c||}{\cellcolor{Gray} \textbf{Deterministic (DET)} } & \multicolumn{3}{c||}{\cellcolor{Gray}  \textbf{Partially Stochastic (PS)} } & \multicolumn{3}{c||}{\cellcolor{Gray} \textbf{Fully Stochastic (FS)} }  \\
 \hline
 \hline
\textbf{Action} & \textbf{A} & \textbf{B} & \textbf{C} & \textbf{A} & \textbf{B} & \textbf{C} & \textbf{A} & \textbf{B} & \textbf{C} \\ 
 \hline
\textbf{A} & $(.8,.8)$ & $(-1.,-1.)$ & $(0,0)$ & $(.8,.8)$  & $(-1.,-1.)$ & $(0,0)$ & $(.9/.7,.9/.7)$ & $(.2/-1,.2/-1)$ & $(.6/-.6,.6/-.6)$ \\ 
 \hline
\textbf{B} & $(-1.,-1.)$ & $(.6,.6)$ & $(.5,.5)$ & $(-1.,-1.)$ & $(1.0/0, 1.0/0)$* & $(.5,.5)$ & $(.2/-1,.2/-1)$ & $(1.0/0,1.0/0)$* & $(.9/.1,.9/.1)$ \\  \hline
\textbf{C} & $(.0,.0)$ & $(.0,.0)$ & $(.4,.4)$ & $(.0,.0)$ & $(.0,.0)$ & $(.4,.4)$ & $(.6/-.6,.6/-.6)$ & $(.4/-.4,.4/-.4)$ &  $(.8/0,.8/0)$ \\  \hline
\end{tabular}}
\caption{Reward structures for Deterministic (DET), Partially Stochastic (PS) and Fully Stochastic (FS) Apprentice Firemen Games, to be interpreted as rewards for (Agent 1, Agent 2). For (B,B)* within PS and FS 1.0 is yielded on 60\% of occasions.}
\label{fig:Apprentice_Firemen_Games_Rewards}
\end{figure*}

\section{Empirical Evaluation} 

\subsection{MA-DRL implementation details}

Given the number of cells within the main area that agents can occupy (90), fire locations (25), agent color combinations (16) and actions (5) we estimate 16,020,000 state-action pairs per layout before factoring in civilians and additional layout specific cells. We therefore follow the example of recent publications by conducting evaluations in gridworlds with sufficient complexity to warrant a MA-DRL approach  \cite{leibo2017multi, guptacooperative, palmer2018lenient}. Networks consist of 2 convolutional layers with 32 and 64 kernels respectively, a fully connected layer (1024 neurons) and an output node for each action\footnote{We make our code available online: \url{https://github.com/gjp1203/nui_in_madrl}}. We use learning rate $\alpha=0.0001$, discount rate $\gamma=0.95$ and $\epsilon$-Greedy exploration with a $\epsilon$ decay rate of 0.999. Each $ERM$ stores 250,000 transitions. Regarding algorithm specific configurations:

\noindent{\bf NUI-DDQN.} To determine the set of trajectories that $\tau$ belongs to each oracle $\vartheta$ queries the respective agent instance regarding the ID of the equipment used. Each $ERM_u$ stores 100 episodes, while the decay rate for $r_u^{min}$ is set to 0.995.

\noindent{\bf LDDQNs.} We use a leniency moderation factor $K=1.0$ and a \emph{retroactive temperature decay schedule} combined with \emph{temperature greedy exploration} strategy as described in \cite{palmer2018lenient}.

\noindent{\bf HDDQNs.} In our evaluation below $\beta$ is the decimal portion of $\alpha$. 

\subsection{Experiments}

To evaluate our hypothesis in Section \ref{sec:AFG} we collect 30 training runs of 5,000 episodes per algorithm within each layout. For LDDQNs an additional 5,000 episodes are required to sufficiently decay the temperatures $\mathcal{T}(o_i,a_i)$. Finally, to evaluate the impact of stochastic transitions we introduce 10 civilians in layout 2 and conduct 30 runs of 10,000 episodes per setting. 

\subsection{Evaluation using phase plots}

The ternary phase plots depicted in Table \ref{tbl:learning_dynamics_0C} provide insights regarding the learning dynamics of the agents. Each line illustrates the average shift in the trajectory distributions throughout the runs conducted, using a rolling window of 1000 episodes. The black squares at the centre of each plot represent the averaged initial $T^u$ distributions while the red dots represents the final distribution. Each corner represents 100\% of trajectories $\tau \in T^u$ for the labelled action~$u~\in~U$. For example, if both lines end with red dots in the top corner of a simplex, then the two agents are predominately producing trajectories $\tau \in T^A$, and have converged upon the optimal Nash Equilibrium $(A, A)$. The agents have therefore learned policies where the optimal equipment is being selected from the pickup area in the AFG, as outlined in section \ref{sec:AFG}.    

\noindent{\bf Determistic rewards.} Under pathologies (Section \ref{sec:Pathologies}) we discuss how maximum based learners can overcome relative overgeneralisation in the deterministic reward Climb Game. Similarly the phase plots for deterministic reward settings confirm that with sufficient optimism / leniency, ILs can overcome relative overgeneralisation while facing the curse of dimensionality (HDDQN $\beta=0.5$, LDDQN and NUI-DDQN in Table \ref{tbl:learning_dynamics_0C}). Meanwhile, HDDQN $(\beta=0.9)$ shows that agents with insufficient optimism gravitate towards the \emph{shadow equilibrium} $(C, C)$. Interestingly in layout 2 with 10 civilians HDDQN $(\beta=0.9)$ agents are completing the climb steps discussed in Section \ref{sec:relative_overgeneralisation} towards $(B,B)$, while $\beta=\{0.5, 0.7\}$ converge towards superior joint policies compared to layout 2 without civilians. Further investigation is required to establish why.

\smallskip

\noindent{\bf Stochastic Rewards.} As evident by the phase plots in Table \ref{tbl:learning_dynamics_0C} the optimism that helps HDDQNs overcome relative overgeneralisation in the deterministic reward settings can lead agents to converge upon sub-optimal joint policies when learning from PS or FS rewards. For HDDQN $(\beta=0.5)$ for instance we observe an increase in $\tau \in T^B$ for PS, and $\tau \in T^C$ by agent 2 for FS rewards. LDDQNs meanwhile are less vulnerable, gravitating towards optimal joint-policies despite stochastic rewards and relative overgeneralisation in layout 1 and when receiving PS rewards in layout 2 with no civilians. However, LDDQNs struggle when receiving FS rewards in layout 2, and have limited success once civilians are added. NUI-DDQNs meanwhile predominately converge upon optimal joint-policies. When receiving PS rewards NUI-DDQNs are initially tempted by the misleading rewards received for $(B, B)$, before converging on a joint-policy with the majority of trajectories $\tau \in T^A$. For FS rewards a slight increase in $\tau \in T^C$ can be observed.

\begin{table*}
    \centering
    \resizebox{\linewidth}{!}{%
    \bgroup
    \footnotesize
    \begin{tabu}{|c|c|c|c|c|c|c|c|c|}
        \hline     
            \multicolumn{3}{|c|}{\cellcolor{Gray} \textbf{Layout 1 (Civilians: 0, Episodes: 5,000)} } & \multicolumn{3}{c|}{\cellcolor{Gray}  \textbf{Layout 2 (Civilians: 0, Episodes: 5,000)} } & \multicolumn{3}{c|}{\cellcolor{Gray} \textbf{Layout 2 (Civilians: 10, Episodes: 10,000)} }  \\
        \hline
        \hline
        \rowcolor{MediumGray}
        \textbf{DET} & \textbf{PS} & \textbf{FS} & \textbf{DET} & \textbf{PS}  & \textbf{FS} & \textbf{DET} & \textbf{PS}  & \textbf{FS} \\
        \hline
        \hline
            \multicolumn{9}{|c|}{\cellcolor{LightGray} \textbf{Hysteretic-DDQN $\beta$=0.5}}  \\
        \hline
        \raisebox{-.5\height}{\includegraphics[height=1.7cm]{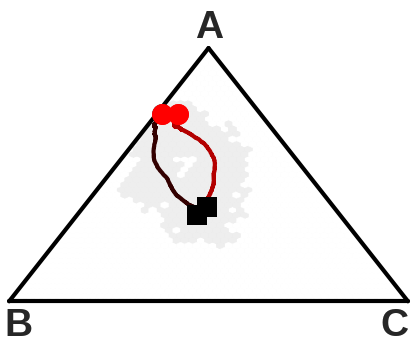}}
        & 
        \raisebox{-.5\height}{\includegraphics[height=1.7cm]{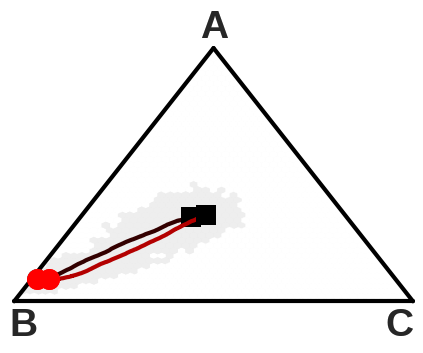}}
        & 
        \raisebox{-.5\height}{\includegraphics[height=1.7cm]{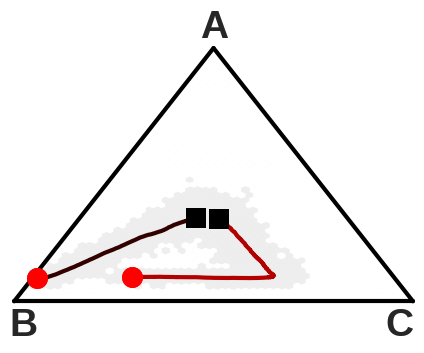}}
        & 
        \raisebox{-.5\height}{\includegraphics[height=1.7cm]{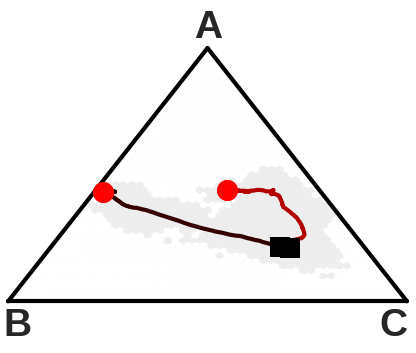}}
        & 
        \raisebox{-.5\height}{\includegraphics[height=1.7cm]{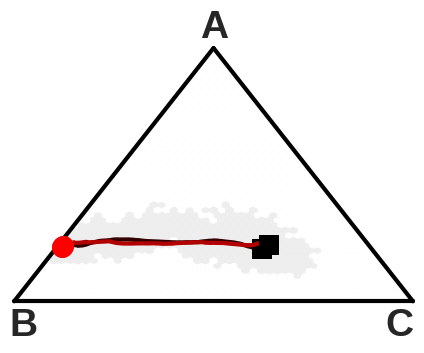}}
        &
        \raisebox{-.5\height}{\includegraphics[height=1.7cm]{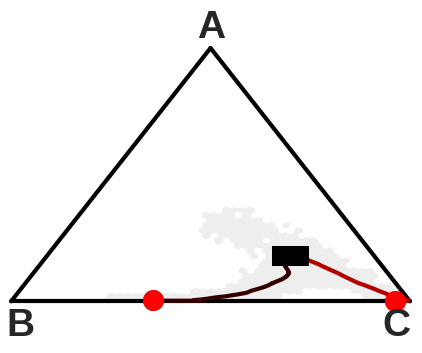}}
        & 
        \raisebox{-.5\height}{\includegraphics[height=1.7cm]{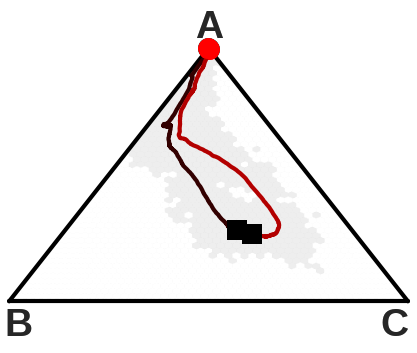}}
        & 
        \raisebox{-.5\height}{\includegraphics[height=1.7cm]{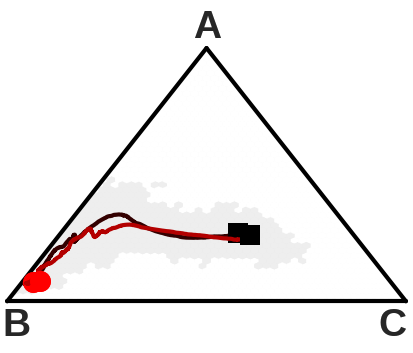}}
        &
        \raisebox{-.5\height}{\includegraphics[height=1.7cm]{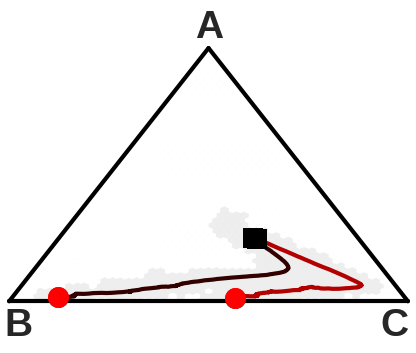}}
        \\
        \hline
        \hline
            \multicolumn{9}{|c|}{\cellcolor{LightGray} \textbf{Hysteretic-DDQN $\beta$=0.7}}  \\
        \hline
        \raisebox{-.5\height}{\includegraphics[height=1.7cm]{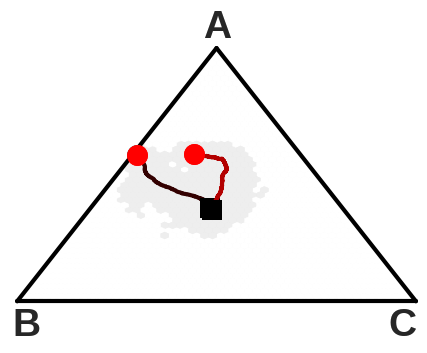}}
        & 
        \raisebox{-.5\height}{\includegraphics[height=1.7cm]{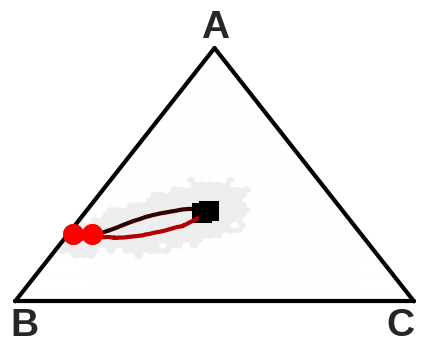}}
        & 
        \raisebox{-.5\height}{\includegraphics[height=1.7cm]{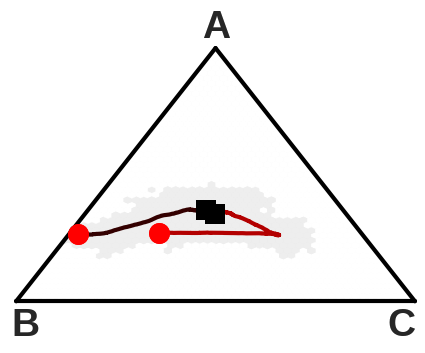}}
        & 
        \raisebox{-.5\height}{\includegraphics[height=1.7cm]{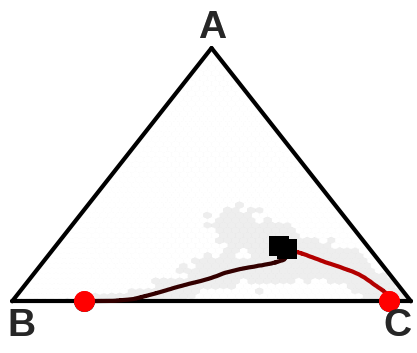}}
        & 
        \raisebox{-.5\height}{\includegraphics[height=1.7cm]{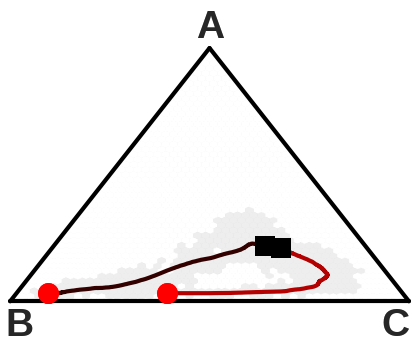}}
        &
        \raisebox{-.5\height}{\includegraphics[height=1.7cm]{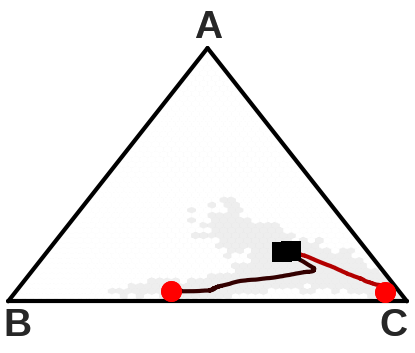}}
        & 
        \raisebox{-.5\height}{\includegraphics[height=1.7cm]{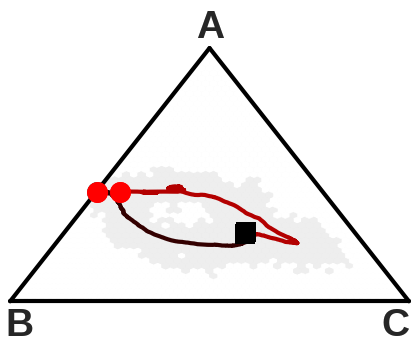}}
        & 
        \raisebox{-.5\height}{\includegraphics[height=1.7cm]{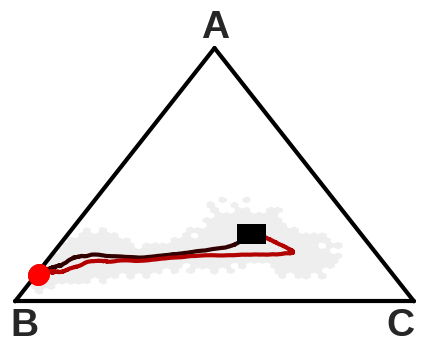}}
        &
        \raisebox{-.5\height}{\includegraphics[height=1.7cm]{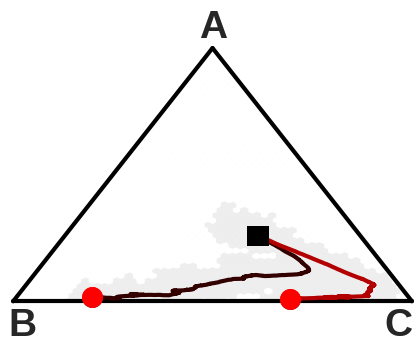}}
        \\
        \hline       
        \hline       
            \multicolumn{9}{|c|}{\cellcolor{LightGray} \textbf{Hysteretic-DDQN $\beta$=0.9}}  \\
        \hline
        \raisebox{-.5\height}{\includegraphics[height=1.7cm]{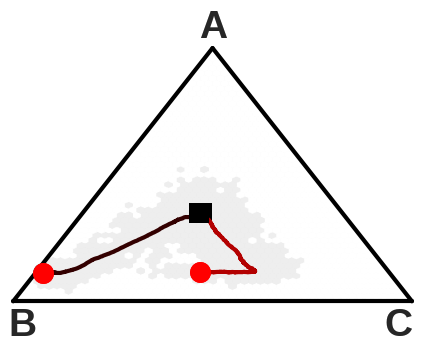}}
        & 
        \raisebox{-.5\height}{\includegraphics[height=1.7cm]{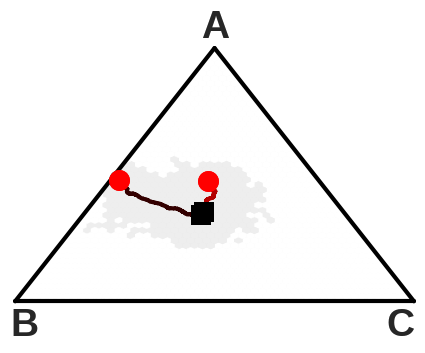}}
        & 
        \raisebox{-.5\height}{\includegraphics[height=1.7cm]{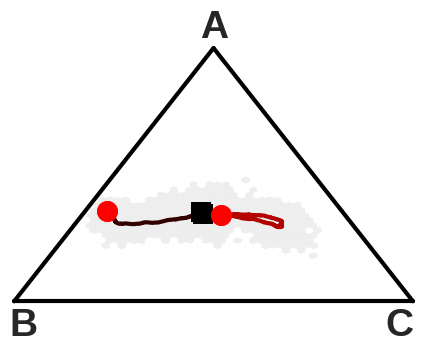}}
        & 
        \raisebox{-.5\height}{\includegraphics[height=1.7cm]{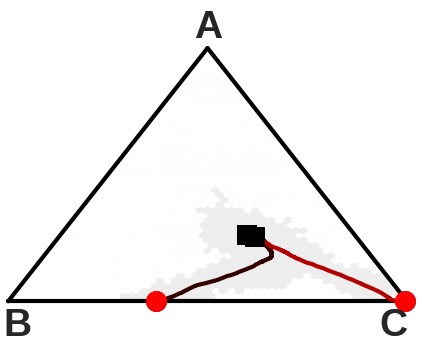}}
        & 
        \raisebox{-.5\height}{\includegraphics[height=1.7cm]{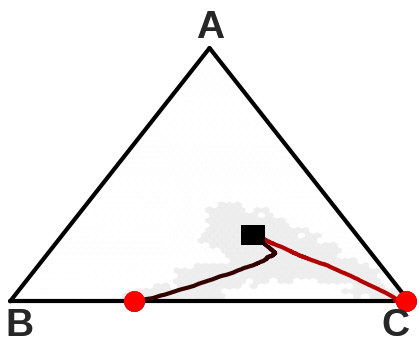}}
        &
        \raisebox{-.5\height}{\includegraphics[height=1.7cm]{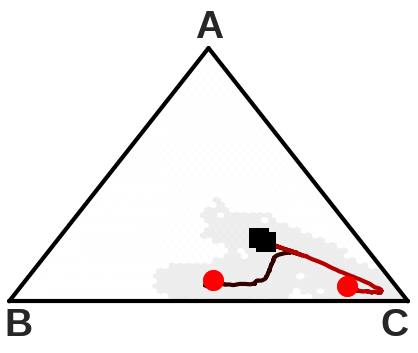}}
        & 
        \raisebox{-.5\height}{\includegraphics[height=1.7cm]{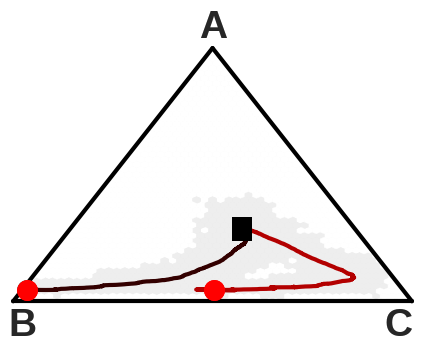}}
        & 
        \raisebox{-.5\height}{\includegraphics[height=1.7cm]{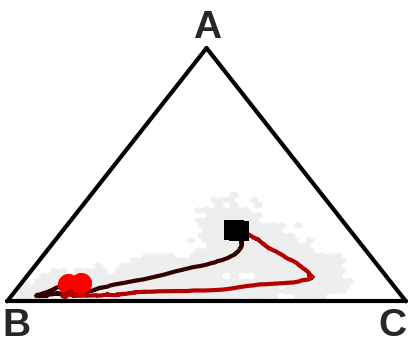}}
        &
        \raisebox{-.5\height}{\includegraphics[height=1.7cm]{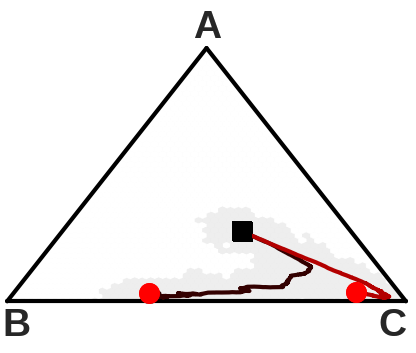}}
        \\
        \hline
        \hline
            \multicolumn{9}{|c|}{\cellcolor{LightGray} \textbf{Lenient-DDQN}}  \\

        \hline       
        \raisebox{-.5\height}{\includegraphics[height=1.7cm]{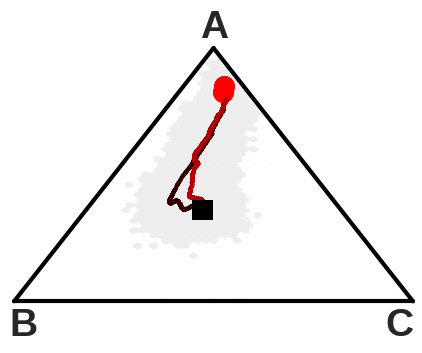}}
        & 
        \raisebox{-.5\height}{\includegraphics[height=1.7cm]{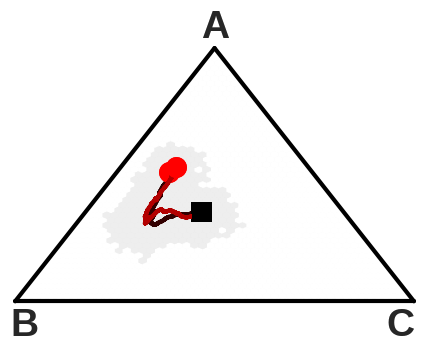}}
        & 
        \raisebox{-.5\height}{\includegraphics[height=1.7cm]{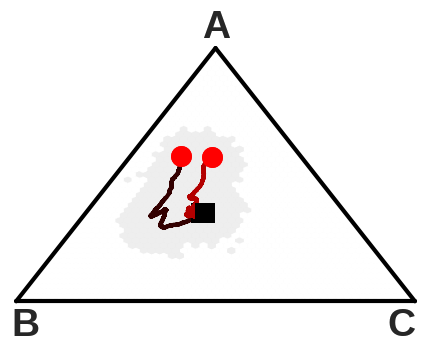}}
        & 
        \raisebox{-.5\height}{\includegraphics[height=1.7cm]{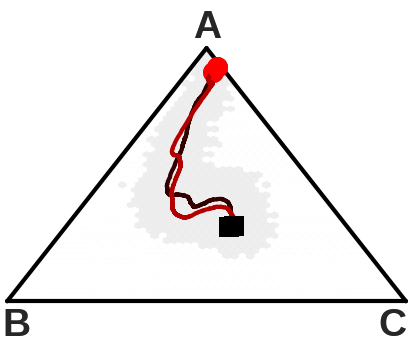}}
        & 
        \raisebox{-.5\height}{\includegraphics[height=1.7cm]{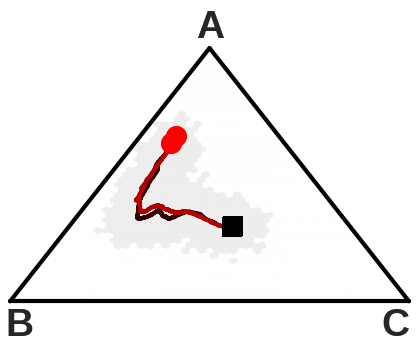}}
        &
        \raisebox{-.5\height}{\includegraphics[height=1.7cm]{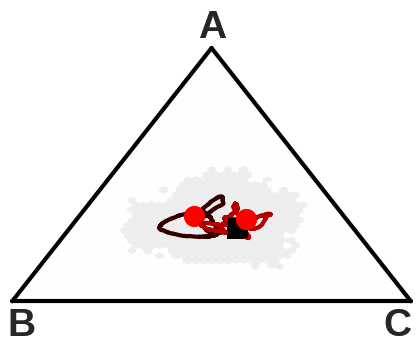}}
        & 
        \raisebox{-.5\height}{\includegraphics[height=1.7cm]{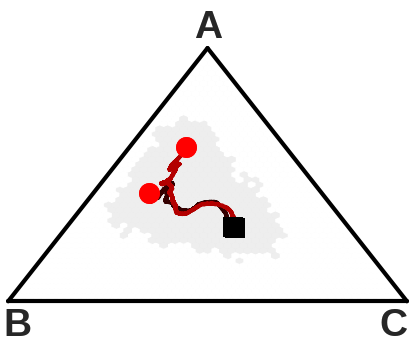}}
        & 
        \raisebox{-.5\height}{\includegraphics[height=1.7cm]{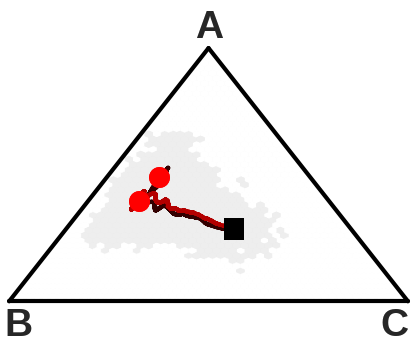}}
        &
        \raisebox{-.5\height}{\includegraphics[height=1.7cm]{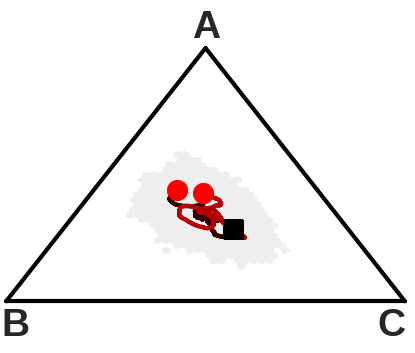}}
        \\          
        \hline       
        \hline
            \multicolumn{9}{|c|}{\cellcolor{LightGray} \textbf{NUI-DDQN}}  \\
        \hline
        \raisebox{-.5\height}{\includegraphics[height=1.7cm]{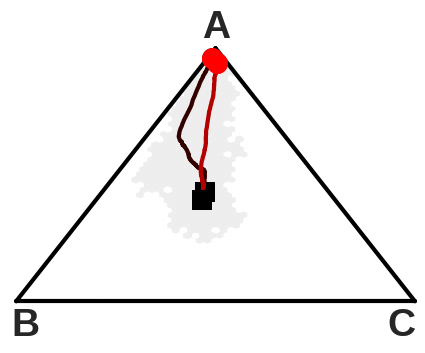}}
        & 
        \raisebox{-.5\height}{\includegraphics[height=1.7cm]{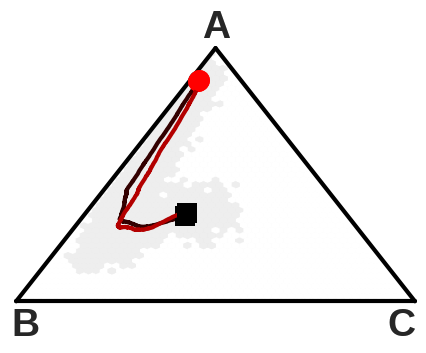}}
        & 
        \raisebox{-.5\height}{\includegraphics[height=1.7cm]{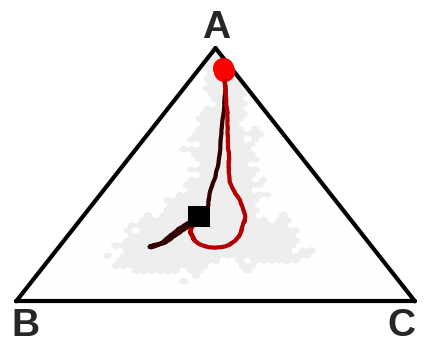}}
        & 
        \raisebox{-.5\height}{\includegraphics[height=1.7cm]{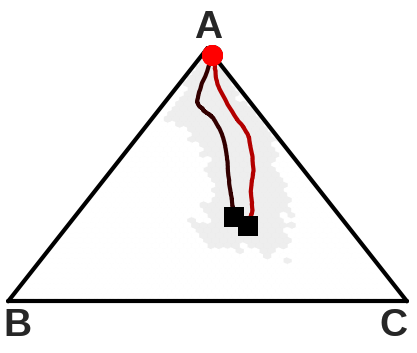}}
        & 
        \raisebox{-.5\height}{\includegraphics[height=1.7cm]{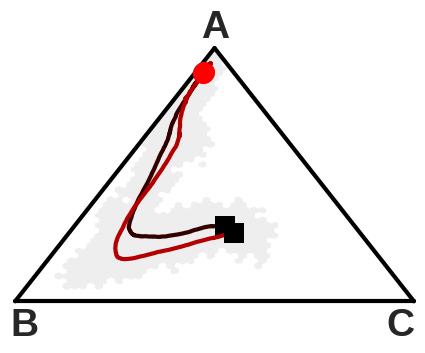}}
        &
        \raisebox{-.5\height}{\includegraphics[height=1.7cm]{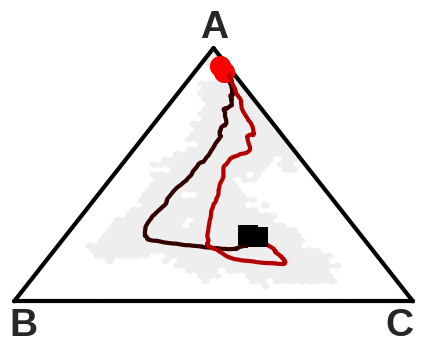}}
        & 
        \raisebox{-.5\height}{\includegraphics[height=1.7cm]{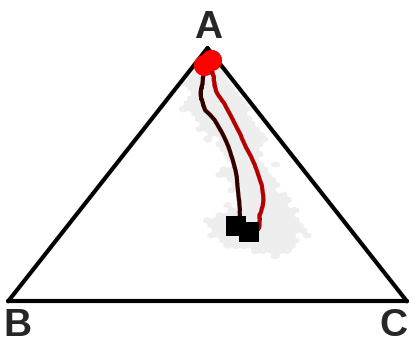}}
        & 
        \raisebox{-.5\height}{\includegraphics[height=1.7cm]{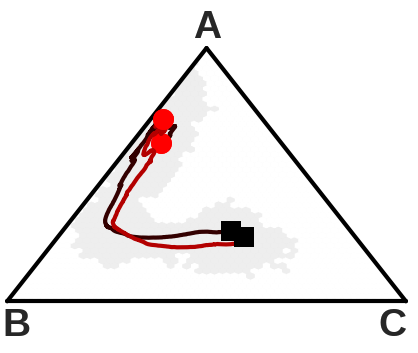}}
        &
        \raisebox{-.5\height}{\includegraphics[height=1.7cm]{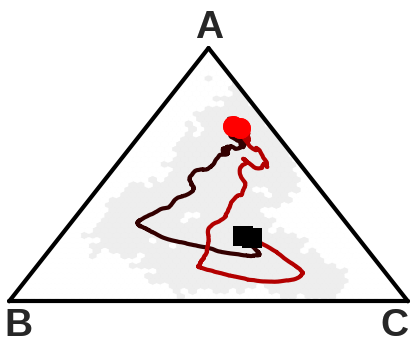}}
        \\
        \hline
        \hline            
        \multicolumn{9}{|c|}{ \includegraphics[width=0.9\columnwidth]{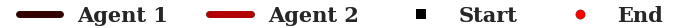}} \\
        \hline
    \end{tabu}\egroup}
    \caption{Phase plots illustrating the average shift in the action $\bm{u}$ distributions throughout the runs conducted, using a rolling window of 1000 episodes. The black squares and red dots represent the initial and final distributions, respectively.}
    \label{tbl:learning_dynamics_0C}
\end{table*}

\subsection{Learning best response policies}

In Section \ref{sec:AFG} we proposed that ILs should learn to avoid miscoordination trajectories within layout 1 due to observing each other during interactions with the equipment pick-up area. To compare the policies learned in layouts 1 and 2 (without civilians), we compute the average coordinated rewards $\overline{RC}$ for each training run. We compute $\overline{RC}$ using the rewards from the final 1000 episodes that did not end in miscoordination outcomes $\{(A, B), (B, A)\}$. Runs with $\overline{RC} \approx 0.8$ have converged upon the optimal joint-policy, where $(A, A)$ is the most frequently observed outcome. For the majority of settings higher $\overline{RC}$ values are achieved by agents in \emph{layout 1}. The scatter plots in Table \ref{tbl:best_responses} provide evidence to support our hypothesis. Each marker within the scatter plots represents the $\overline{RC}$ for an individual run. To provide further clarity we sort the runs by $\overline{RC}$. We observe that HDDQN $\beta=0.7$ and $\beta=0.9$ converges upon a policy with $\overline{RC} \approx 0.8$ numerous times in each reward setting in layout 1, while only twice in layout 2 (HDDQN $\beta=0.7$, PS \& FS)~\footnote{We provide additional $\overline{RC}$ scatter plots for each evaluation setting in the Appendix.}. Interestingly we find that for HDDQN ($\beta=0.5$, PS) and LDDQN (DET \& PS) a larger number of runs converge upon joint-policies where $\overline{RC} \approx 0.8$ in layout 2. NUI-DDQNs meanwhile perform consistently when receiving deterministic and partially stochastic rewards in both settings, while a couple of runs faltered for fully stochastic rewards within layout 2. It is worth noting that even for NUI-DDQN runs with low $\overline{RC}$, $(A, A)$ remains a frequently observed outcome.

\begin{table}[h]
\centering
\footnotesize
\resizebox{\linewidth}{!}{%
\bgroup
\begin{tabu}{|c|c|c|}
\hline
\rowcolor{Gray}
\textbf{Deterministic} &  \textbf{Partially Stochastic} & \textbf{Fully Stochastic} \\
\hline
\hline
\multicolumn{3}{|c|}{\cellcolor{LightGray} \textbf{HDQN $\bm{\beta=0.7}$}} \\
\hline
\includegraphics[height=2.2cm]{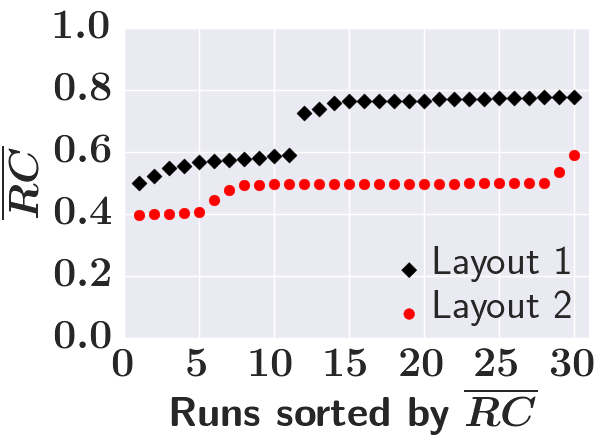}
& 
\includegraphics[height=2.2cm]{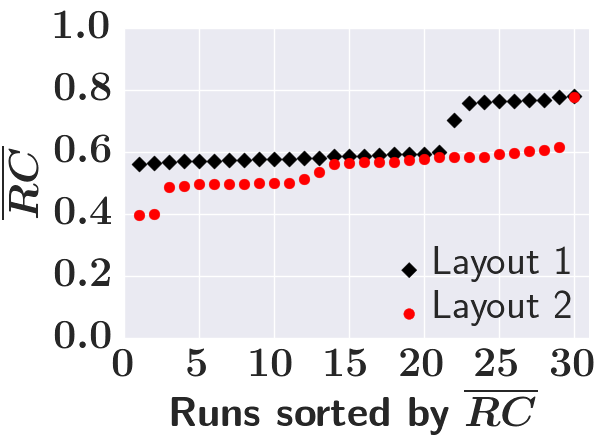}
& 
\includegraphics[height=2.2cm]{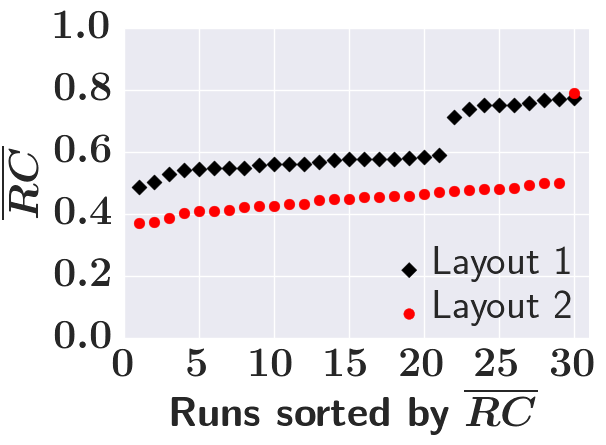} \\
\hline
\hline
\multicolumn{3}{|c|}{\cellcolor{LightGray} \textbf{HDQN $\bm{\beta=0.9}$}} \\
\hline
\includegraphics[height=2.2cm]{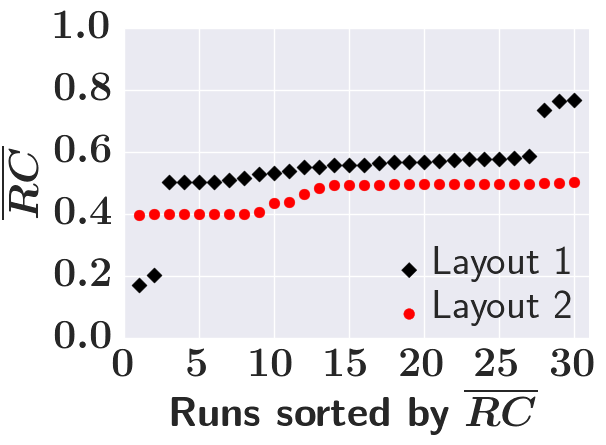}
& 
\includegraphics[height=2.2cm]{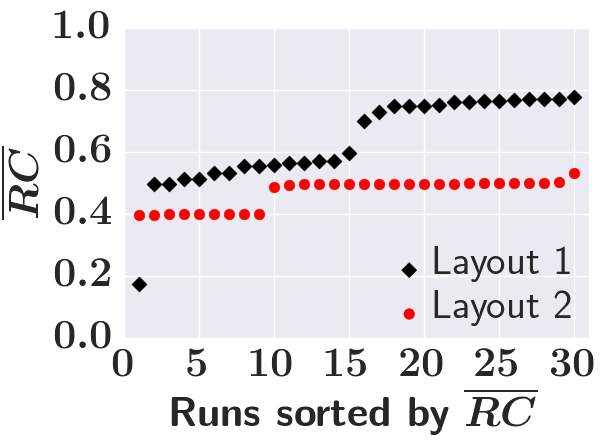}
& 
\includegraphics[height=2.2cm]{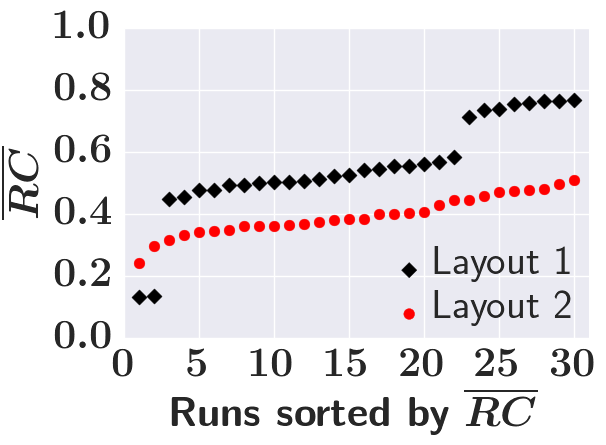} \\
\hline
\end{tabu}\egroup}
\caption{Scatter plots depicting the average coordinated rewards $\bm{\overline{RC}}$ for HDDQNs with $\bm{\beta=0.7}$ and $\bm{\beta=0.9}$.} 
\label{tbl:best_responses}
\end{table}

\subsection{Impact of stochastic transitions} \label{sec:Impact_of_stochastic_transitions}

\begin{figure}[H]
\centering
    \begin{subfigure}[b]{0.49\columnwidth}
        \centering
        \includegraphics[width=\columnwidth]{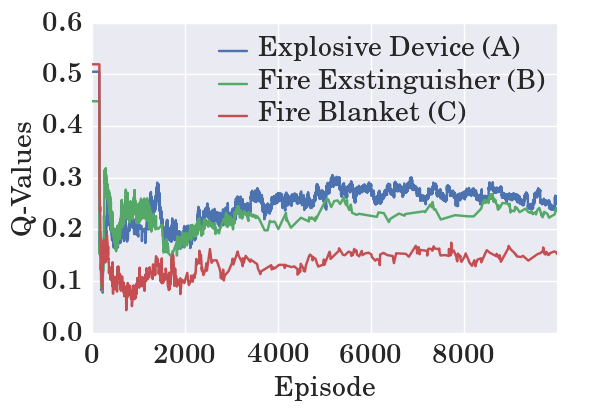} 
        \caption{0 Civilians}
        \label{fig:NUI-DDQN_pickup_qvals:0C}
    \end{subfigure}
    \begin{subfigure}[b]{0.49\columnwidth}
        \centering
        \includegraphics[width=\columnwidth]{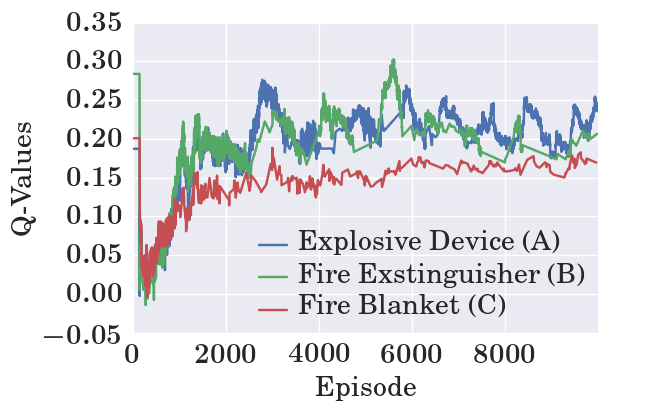}
        \caption{10 Civilians}
        \label{fig:NUI-DDQN_pickup_qvals:10C}
    \end{subfigure}
\caption{NUI-DDQN Pickup Q-Values}
\label{fig:NUI-DDQN_pickup_qvals}
\end{figure}

Introducing 10 civilians to layout 2 allows us to examine the challenges faced by MA-DRL agents when attempting to overcome relative overgeneralisation while making decisions using noisy utility values backed up from stochastic follow-on transitions. In Figure \ref{fig:NUI-DDQN_pickup_qvals} we compare the Q-Values from actions leading to the selection of equipment within both 0 and 10 civilian settings from two individual NUI-DDQN runs with PS rewards. We observe that Q-Values oscillate significantly upon introducing civilians, with Q-Values belonging to sub-optimal equipment $B$ pickups frequently rising above those belonging to $A$. Stochastic transitions can therefore lead to the moving target problem (Section \ref{sec:moving_target}), in this case resulting in extended periods of miscoordination. However, by maintaining negative update intervals, NUI-DDQN can overcome miscoordination and revert back to a policy that generates trajectories $\tau \in T^A$. 

\subsection{Considerations regarding LDDQNs} \label{sec:LDDQN_Considerations}

The phase plots in Table \ref{tbl:learning_dynamics_0C} indicate that given more time an increase in trajectories $\tau \in T^A$ should be possible for LDDQNs. However, while searching for an optimal set of hyperparameters we became aware of one particular dilemma: while LDDQNs fail to converge upon $(A, A)$ with insufficiently decayed temperature values, rapidly decaying temperature values leaves LDDQNs vulnerable during the periods of miscoordination discussed in Section \ref{sec:Impact_of_stochastic_transitions}. We therefore choose a patient approach, with the consequence that even after 10,000 episodes the agents have still not converged. To illustrate this dilemma we conduct additional runs in a simplified PS reward version of layout 1 with only 1 fire location. By varying the number of obstacles surrounding this fire, and thereby controlling the number of \emph{Access Points} from which the agents can extinguish it, we observe the rolling percentage of $(A, A)$ outcomes increases significantly faster when the number of  $\mathcal{T}_t(o_i, a_i)$ values that need decaying decreases. We conduct 20 runs for each \emph{access point} setting:

\begin{figure}[h]
\centering
\includegraphics[width=0.6\columnwidth]{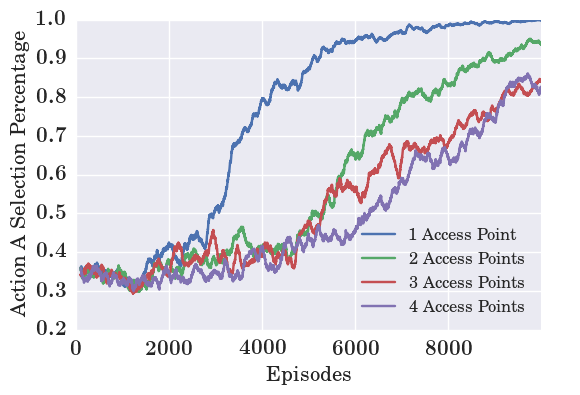} 
\caption{Running $\bm{(A,A)}$ \% by LDDQNs dependent on fire \emph{Access Points}. Agents could overlap next to the fire for \emph{1 Access Point}.}
\label{fig:access_point_impact_on_ldqn}
\end{figure} 

\section{Related Literature}

ILs within MA-DRL systems are receiving an increasing amount of attention within literature. However, a lot of this research focuses on the impact of stochasticity and the amplified moving target problem caused by ERMs \cite{hernandez2018multiagent}. Zheng et al. \cite{zheng2018weighted} introduce a Weighted Double-DQN that makes use of a lenient reward network along with a scheduled replay strategy to improve the convergence rate within stochastic cooperative environments. Foerster et al. \cite{foerster2017stabilising} show that the ERM can be stabilised by using importance sampling as a means to identify outdated state transitions by maintaining observation histories of the actions taken by the other agents in the system. A further active area within MA-DRL research is centralized training for decentralized execution. Gupta et al. \cite{guptacooperative} evaluate policy gradient, temporal difference error, and actor critic methods on cooperative control tasks that include discrete and continuous state and action spaces, using a decentralized parameter sharing approach with centralized learning. Rashid et al. \cite{rashid2018qmix} propose QMIX, introducing a network that estimates joint action-values as a non-linear combination of agent-values, thereby helping agents learn optimal joint-action values based on the additional information that centralized learning provides. Perhaps the most relevant recent work to our research is by Wei et al. \cite{wei2018multiagent}, who introduce Multiagent Soft Q-Learning, an algorithm which converges towards a superior local optima compared to MA-DDPG \cite{lowe2017multi} within continuous action domains suffering from relative overgeneralisation. In contrast to our approach Multiagent Soft Q-Learning augments rewards with an entropy term, thereby increasing the likelihood of finding multiple modes within a continuous action space compared to the discrete action space used in our experiments \cite{wei2018multiagent, haarnoja2017reinforcement}. However, Multiagent Soft Q-Learning is currently a centralized approach that has only been tested within a single state continuous game for two agents, with the authors currently investigating the algorithm's scalability to ILs within sequential continuous games. 

\section{Future work} \label{ref:future_work:improved_oracle}

\noindent{\bf 1)} We conducted trials using 20 sets of fixed policies trained in a simplified layout 1, finding evidence that leader-follower dynamics emerge during training, where one agent waits to observe the other's equipment choice. We assigned the policies to disjoint sets $\Pi^A$ and $\Pi^B$ based on the percentage of $\tau \in T^A$ and $\tau \in T^B$ their roll-outs produced. Upon subsequently running trials with each leader-follower combination, we find half of $\pi^B$ followers choose $A$ when paired with a $\pi^A$ leader and vice-versa. We are currently investigating why only some ILs develop this adaptive ability.

\noindent{\bf 2)} For simplicity, the actions $u \in U$ returned by the oracle $\vartheta$ for the AFG are predefined. Going foward, recent work on the topic of \emph{theory of mind} by Rabinowitz et al. \cite{pmlr-v80-rabinowitz18a} could pave the way for a learned oracle $\vartheta$. The authors build a data-efficient meta-learner that learns models of the agents that it encounters. Through observing trajectories $\tau$ their resulting \emph{theory of mind network} architecture is used to predict next-step actions, the consumption of objects within the environment and the successor representation. This opens up the possibility of applying NUI-DDQN to more complex domains where a learning approach is required to identify actions $u \in U$.

\section{Conclusions} 

Our empirical evaluation highlights the challenges MA-DRL agents must overcome, to avoid converging upon sub-optimal joint policies when making decisions using noisy approximated utility estimates backed-up from stochastic follow-on state-transitions and rewards.\\
\noindent
To summarize our contributions:\\
\noindent{\bf 1)} We presented the Apprentice Firemen Game (AFG), which is a new and challenging environment that simultaneously confronts learners with four pathologies: relative overgeneralisation, stochasticity, the moving target problem, and alter exploration problem. \\
\noindent{\bf 2)} We evaluate \emph{hysteretic} \cite{omidshafiei2017deep} and \emph{lenient} \cite{palmer2018lenient} learners on the AFG. While both approaches can overcome the pathologies in simpler settings, they fail when required to independently make irrevocable decisions in seclusion determining an episode's outcome.\\
\noindent{\bf 3)} Motivated by this we designed a new algorithm NUI-DDQN that is based on negative update intervals. Our algorithm identifies and discards episodes that end in miscoordination. In doing so, it reduces the noise introduced by the large punishments that result from miscoordination. We show that NUI-DDQN consistently converges towards the optimal joint-policy within each setting. 


\section{Acknowledgments}

We thank the HAL Allergy Group for partially funding the PhD of Gregory Palmer and gratefully acknowledge the support of NVIDIA Corporation with the donation of the Titan X Pascal GPU that enabled this research.


\bibliographystyle{ACM-Reference-Format}  
\balance  

\newpage
\onecolumn
\appendix
\section{Appendices}

\subsection{Hyperparameters}

Table \ref{fig:appendix:hyperparameters} lists the hyperparameters used for our empirical evaluation. To reduce the time required to evaluate LDDQN we apply python's xxhash to masked observations (i.e., removing civilians).

\begin{table}[ht]
\begin{center}
{\small 
\resizebox{0.5\columnwidth}{!}{\begin{tabular}{ | c | c | c |}
\hline
\textbf{Component} & \textbf{Hyperparameter} & \textbf{Range of values} \\
\hline
\hline
\multirow{4}{*}{\textbf{DDQN Base}} & Learning rate $\alpha$ & 0.0001 \\
\cline{2-3}   
& Discount rate $\gamma$ & 0.95 \\
\cline{2-3}   
& Target network sync. steps & 5000  \\
\cline{2-3}   
& ERM Size &  250'000 \\
\hline
\hline
\multirow{ 3}{*}{\textbf{$\epsilon$-Greedy Exploration}} & Initial $\epsilon$ value & 1.0 \\
\cline{2-3}   
& $\epsilon$ Decay factor & 0.999 \\
\cline{2-3}   
& Minimum $\epsilon$ Value & 0.05\\
\hline
\hline
\multirow{8}{*}{\textbf{Leniency}} & MaxTemperature & 1.0 \\
\cline{2-3}   
& Leniency Modification Coefficient $K$ & 1.0 \\
\cline{2-3}   
& TDS Exponent $\rho$ & -0.1 \\
\cline{2-3}   
& TDS Exponent Decay Rate $d$ & 0.95\\
\cline{2-3}   
& Initial Max Temperature Value $\nu$ & 1.0 \\
\cline{2-3}
& Max Temperature Decay Coefficient $\mu$ & 0.9998 \\
\cline{2-3}
& Action Selection Exponent & 0.25 \\
\cline{2-3}
& Hashing & xxhash \\
\hline  
\hline
\multirow{2}{*}{\textbf{NUI-DDQN}} & $ERM_u$ Capacity & 100 Episodes \\
\cline{2-3}   
& Decay threshold & 50 Episodes \\
\cline{2-3}   
& $r_u^{min}$ decay rate & 0.995 \\
\hline  
\end{tabular}}}
\end{center}
\captionsetup{justification=centering}
\caption{Hyper-parameters}
\label{fig:appendix:hyperparameters}
\end{table}   

\subsection{Learning Best Response Policies}

Table \ref{tbl:appendix:avg_coord_reward} provides additional $\overline{RC}$ scatter plots for each evaluation setting. Each marker within the scatter plots represents the $\overline{RC}$ for an individual run. To provide further clarity we sort the runs by $\overline{RC}$. We observe that for the majority of settings higher $\overline{RC}$ values are achieved by agents in \emph{layout 1}. Interestingly only HDDQN ($\beta=0.5$, PS Rewards) and LDDQN (DET \& PS Rewards) achieved higher $\overline{RC}$ values in layout 2. NUI-DDQNs meanwhile perform consistently when receiving deterministic and PS rewards, while a couple of runs faltered for FS rewards within layout 2. It is worth noting that even for NUI-DDQN runs with low $\overline{RC}$, $(A, A)$ remains a frequently observed outcome:

\begin{table}[H]
    \centering
    \resizebox{\textwidth}{!}{%
     \small
    \begin{tabular}{|c|c|c|c|c|c|}
       \cline{1-5}
        \textbf{HDQN ($\bm{\beta=0.5}$)} & \textbf{HDQN ($\bm{\beta=0.7}$)} & \textbf{HDQN ($\bm{\beta=0.9}$)} & \textbf{LDDQN}  &  \textbf{NUI-DDQN} \\
            \hline    
            \raisebox{-.5\height}{\includegraphics[width=3cm]{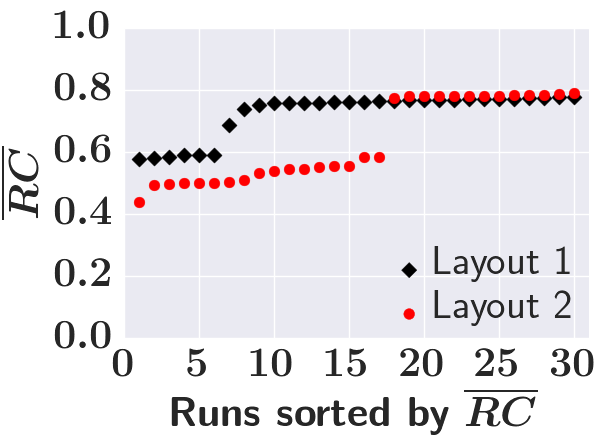}}
            & 
            \raisebox{-.5\height}{\includegraphics[width=3cm]{SCATTER/HDQN07_DET.png}}
            & 
            \raisebox{-.5\height}{\includegraphics[width=3cm]{SCATTER/HDQN09_DET.png}}
            & 
            \raisebox{-.5\height}{\includegraphics[width=3cm]{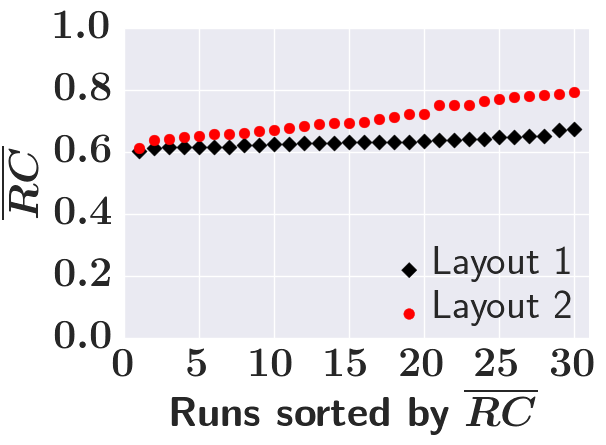}}
            &
            \raisebox{-.5\height}{\includegraphics[width=3cm]{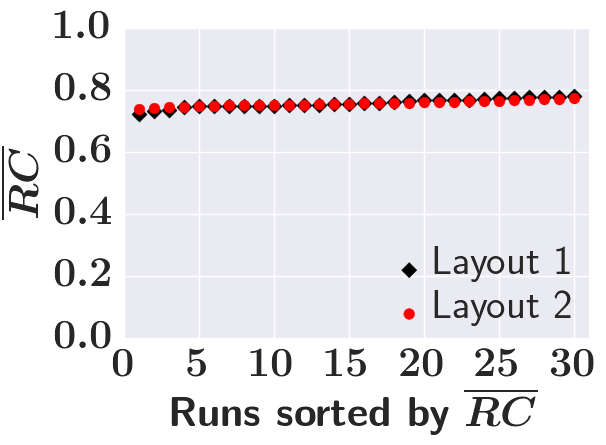}}
            & \textbf{DET}
            \\
            \hline    
            \raisebox{-.5\height}{\includegraphics[width=3cm]{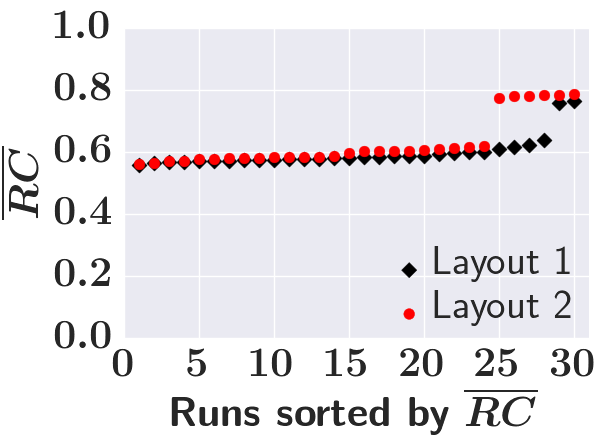}}
            & 
            \raisebox{-.5\height}{\includegraphics[width=3cm]{SCATTER/HDQN07_PS.png}}
            & 
            \raisebox{-.5\height}{\includegraphics[width=3cm]{SCATTER/HDQN09_PS.png}}
            & 
            \raisebox{-.5\height}{\includegraphics[width=3cm]{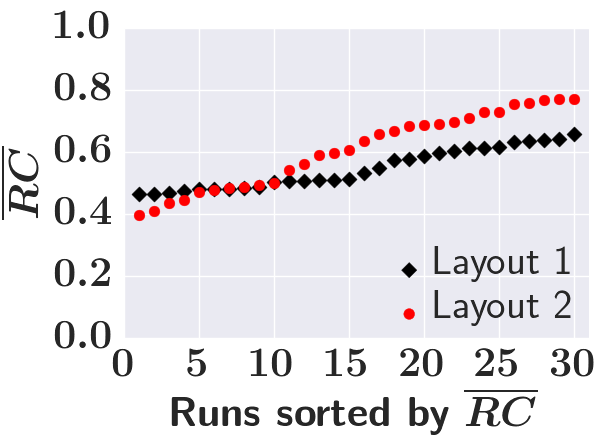}}
            &
            \raisebox{-.5\height}{\includegraphics[width=3cm]{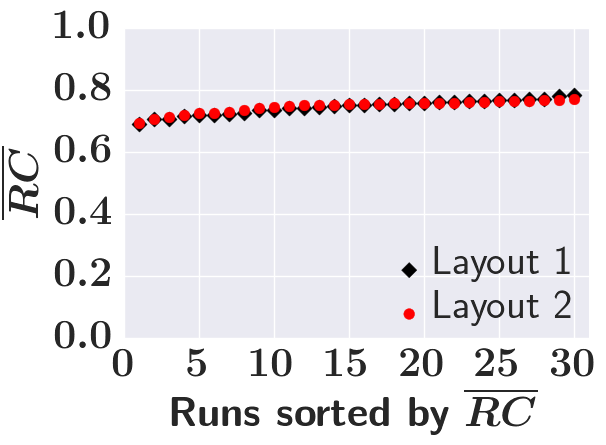}}
            & \textbf{PS}
            \\
        \hline     
              \raisebox{-.5\height}{\includegraphics[width=3cm]{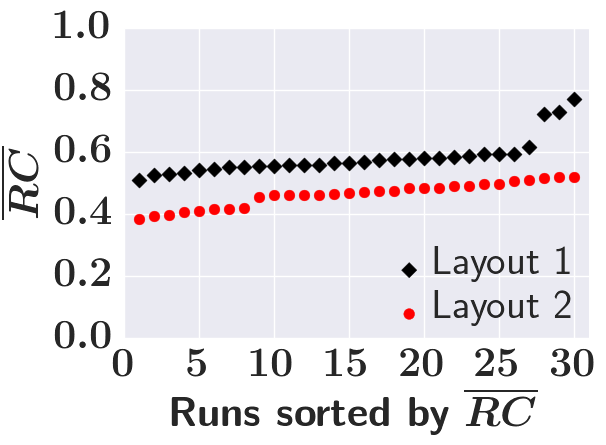}}
            & 
            \raisebox{-.5\height}{\includegraphics[width=3cm]{SCATTER/HDQN07_FS.png}}
            & 
            \raisebox{-.5\height}{\includegraphics[width=3cm]{SCATTER/HDQN09_FS.png}}
            & 
            \raisebox{-.5\height}{\includegraphics[width=3cm]{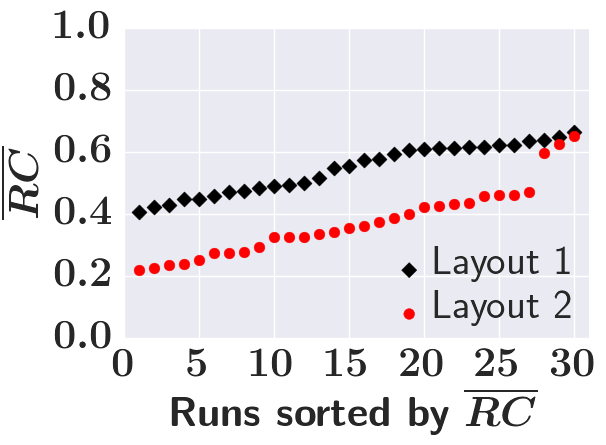}}
            &
            \raisebox{-.5\height}{\includegraphics[width=3cm]{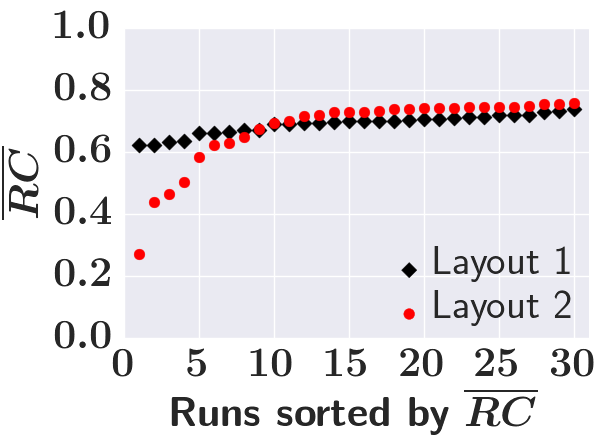}}
            
             & \textbf{FS}
            \\
        \hline          
    \end{tabular}}
    \caption{Scatter plots illustrating the average coordinated reward $\bm{\overline{RC}}$ for each training run. The x-axis is sorted by $\bm{\overline{RC}}$ values.}
    \label{tbl:appendix:avg_coord_reward}
\end{table}

\subsection{LDDQN Variable Access Points Experiments: Layouts \& Phase Plots}

\begin{figure}[h]
\centering
\begin{subfigure}[b]{0.24\columnwidth}
\centering
\includegraphics[width=0.8\columnwidth]{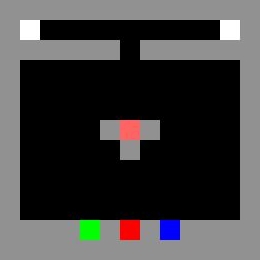}
\caption{1 Access Point}
\end{subfigure}
\centering
\begin{subfigure}[b]{0.24\columnwidth}
\includegraphics[width=0.8\columnwidth]{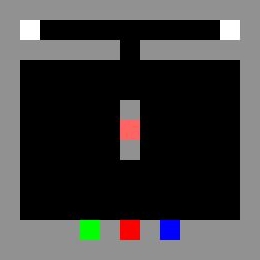}
\caption{2 Access Points}
\end{subfigure}
\centering
\begin{subfigure}[b]{0.24\columnwidth}
\includegraphics[width=0.8\columnwidth]{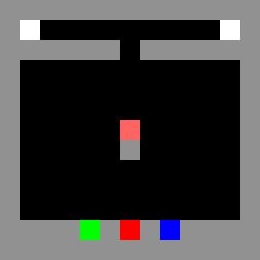}
\caption{3 Access Points}
\end{subfigure}
\centering
\begin{subfigure}[b]{0.24\columnwidth}
\includegraphics[width=0.8\columnwidth]{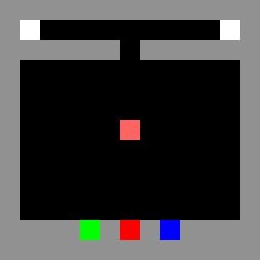}
\caption{4 Access Points}
\end{subfigure}
\begin{subfigure}[b]{0.24\columnwidth}
\centering
\includegraphics[width=0.9\columnwidth]{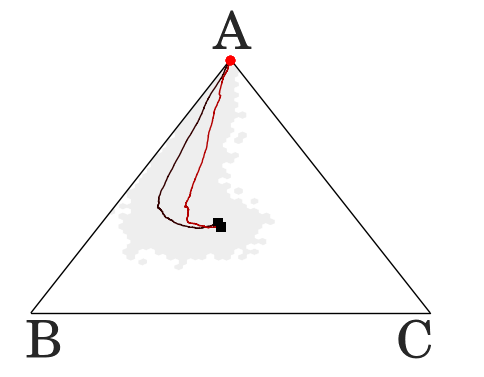}
\caption{1 Access Point}
\end{subfigure}
\centering
\begin{subfigure}[b]{0.24\columnwidth}
\includegraphics[width=0.9\columnwidth]{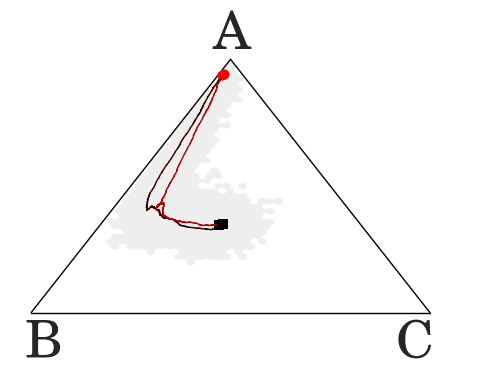}
\caption{2 Access Points}
\end{subfigure}
\centering
\begin{subfigure}[b]{0.24\columnwidth}
\includegraphics[width=0.9\columnwidth]{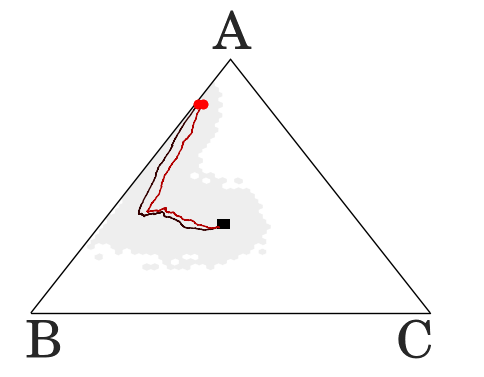}
\caption{3 Access Points}
\end{subfigure}
\centering
\begin{subfigure}[b]{0.24\columnwidth}
\includegraphics[width=0.9\columnwidth]{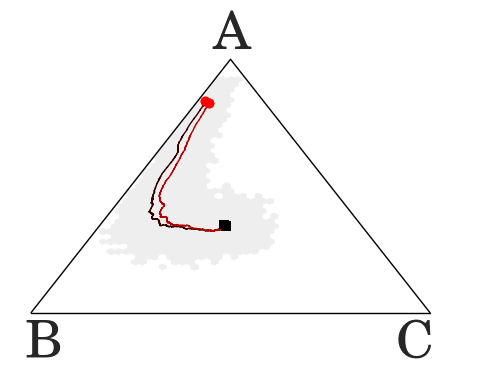}
\caption{4 Access Points}
\end{subfigure}
\caption{Phase plots illustrate delayed convergence of LDDQNs as a result\\ of increasing the number of possible state-action pairs.}
\label{fig:appendix:ap_layouts_and_phase_plots}
\end{figure}

Figure \ref{fig:appendix:ap_layouts_and_phase_plots} illustrates the AFG layouts used for the evaluations discussed in Section \ref{sec:LDDQN_Considerations}. We also provide the resulting phase plot for each layout.

\end{document}